\documentclass[aps, showkeys]{revtex4-2}
\usepackage{graphicx, xcolor, amssymb, amsmath, amsfonts}\usepackage[colorlinks=true, linkcolor=blue, citecolor=blue, urlcolor=blue]{hyperref}
\usepackage{url, soul, ulem, textgreek, bm}
\usepackage[draft]{changes}
\definechangesauthor[name = taegeun, color = blue]{TS}

\begin{document}

\title{Deep learning methods for Hamiltonian parameter estimation and magnetic domain image generation in twisted van der Waals magnets}

\author{Woo Seok Lee}
\affiliation{1ST Biotherapeutics, Yongin 16942, Republic of Korea}


\author{Taegeun Song}
\affiliation{Department of Data Information and Physics, Kongju National University, Kongju 32588, Republic of Korea}
\affiliation{Institute of Application and Fusion for Light, Kongju National University, Cheonan 31080, Republic of Korea}

\author{Kyoung-Min Kim}
\affiliation{Center for Theoretical Physics of Complex Systems, Institute for Basic Science, Daejeon 34126, Republic of Korea}
\email{kmkim@ibs.re.kr}

\begin{abstract}
The application of twist engineering in van der Waals magnets has opened new frontiers in the field of two-dimensional magnetism, yielding distinctive magnetic domain structures. Despite the introduction of numerous theoretical methods, limitations persist in terms of accuracy or efficiency due to the complex nature of the magnetic Hamiltonians pertinent to these systems. In this study, we introduce a deep-learning approach to tackle these challenges. Utilizing customized, fully connected networks, we develop two deep-neural-network kernels that facilitate efficient and reliable analysis of twisted van der Waals magnets. Our regression model is adept at estimating the magnetic Hamiltonian parameters of twisted bilayer CrI\textsubscript{3} from its magnetic domain images generated through atomistic spin simulations. The ``generative model" excels in producing precise magnetic domain images from the provided magnetic parameters. The trained networks for these models undergo thorough validation, including statistical error analysis and assessment of robustness against noisy injections. These advancements not only extend the applicability of deep-learning methods to twisted van der Waals magnets but also streamline future investigations into these captivating yet poorly understood systems.
\end{abstract}


\maketitle
\tableofcontents

\section{Introduction}

Recent discoveries in the field of twist engineering of two-dimensional (2D) magnetism have unveiled unique magnetic domain structures, including nanoscale magnetic domain arrays \cite{Song2021, Xu2022, Xie2022, Xie2023, Cheng2023}. The qualitative features of these structures find their explanation in the context of exotic interlayer magnetic coupling induced by moiré patterns \cite{Hejazi10721, Akram2021, Zheng2022, Kim2023, Yang2023}. However, due to the complex nature of the magnetic coupling in the systems \cite{Kim2023}, the quantitative description and analysis of these structures pose formidable challenges within conventional theoretical frameworks. Specifically, effective model approximations, such as the continuum model \cite{Hejazi10721, PhysRevB.104.L100406, PhysRevB.108.174440} and the free-layer-substrate model \cite{Tong2018, PhysRevB.103.L140406, PhysRevB.104.014410, PhysRevResearch.3.013027, Akram2021, Ghader2022, Fumega_2023}, exhibit substantial limitations in accuracy. Alternative atomistic spin simulations may offer improved accuracy, but they are hindered by their resource-intensive nature \cite{Zheng2022, Kim2023, Yang2023, PhysRevB.108.L100401, Kim2024, Akram2024}. Consequently, it is crucial to develop theoretical frameworks that are both efficient and reliable to overcome these constraints and propel advancements in this field.

In this study, we introduce a deep-learning approach that facilitates efficient and reliable analysis of twisted van der Waals magnets. Deep learning has showcased remarkable success in tackling scientific challenges across various physical systems \cite{doi:10.1126/sciadv.abb0872, Wright2022, SCHMIDHUBER201585, PhysRevResearch.2.033429, https://doi.org/10.1002/prot.25834, doi:10.1126/science.aag2302, PhysRevB.97.035116, PhysRevB.99.174426, PhysRevB.99.024423, Lee2020, Lee2021_1, Lee2021_2, Lee_SF_2020, EASAW2023102053, Miyazaki_2023}. Specifically, the application of deep neural network (DNN) techniques to 2D magnetic systems has proven highly effective in extracting magnetic Hamiltonian parameters from magnetic domain images \cite{doi:10.1126/sciadv.abb0872}. Furthermore, these techniques have demonstrated applicability to experimentally observed domain images within the same research. These results validate DNN techniques as valuable tools for comprehending 2D magnetic domain structures.

\begin{figure}[ht!]
    \centering
    \includegraphics[width=.50\textwidth]{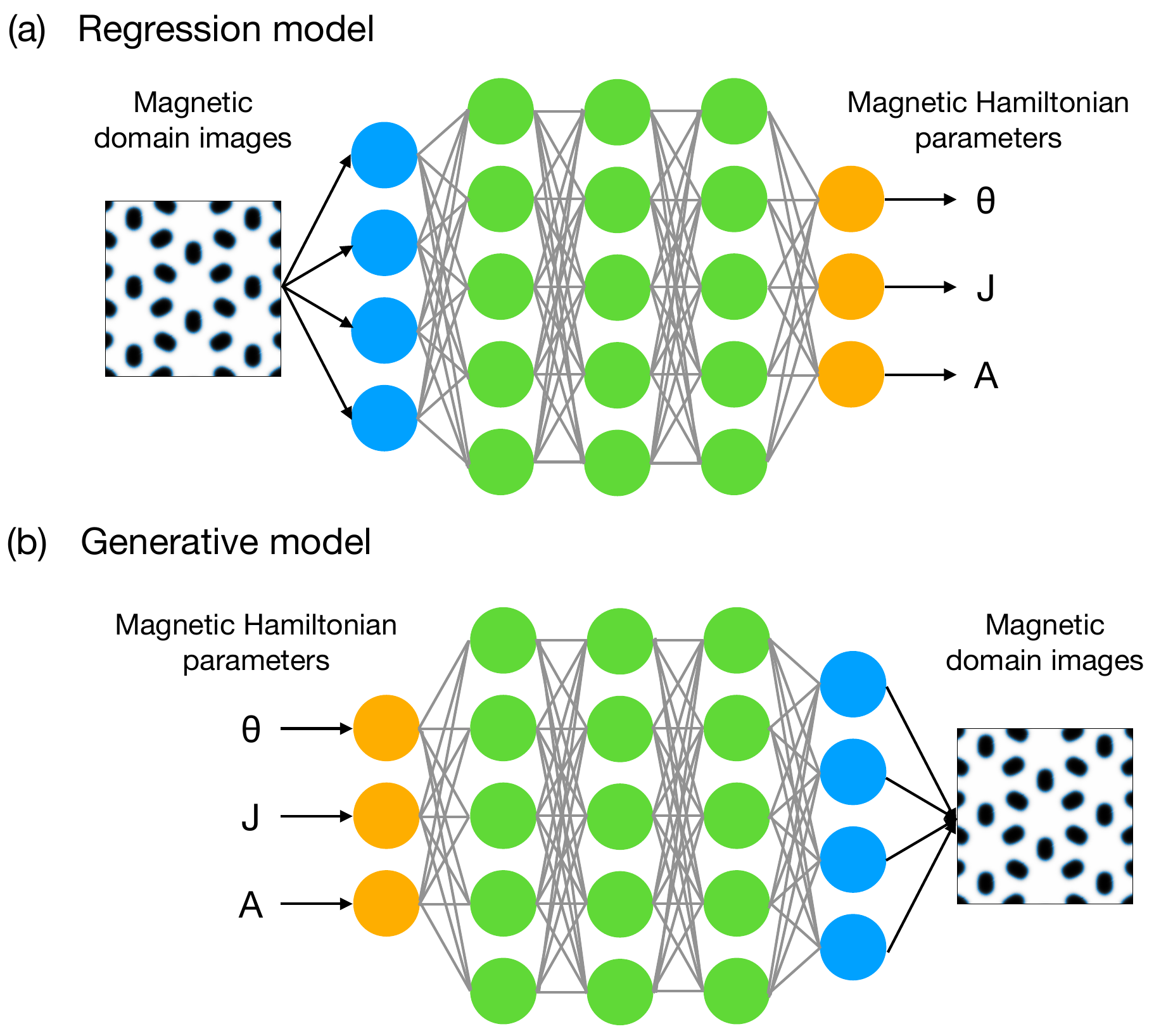}
    \caption[Schematic diagram]{Schematic diagrams illustrating two deep neural network models developed in this study. (a) The regression model for estimating magnetic Hamiltonian parameters from input magnetic domain images. (b) The generative model for producing predicted magnetic domain images based on input parameters.}
    \label{fig1}
\end{figure}

We develop two specialized DNN kernels designed to analyze 2D magnetic domain structures within twisted van der Waals magnets, such as nanoscale magnetic domain arrays found in chromium triiodide (CrI\textsubscript{3}) \cite{Song2021, Xie2022, Xu2022, Xie2023, Cheng2023, Hejazi10721, PhysRevResearch.3.013027, Akram2021, Ghader2022, Zheng2022, Fumega_2023, Kim2023, Yang2023, PhysRevB.108.174440, PhysRevB.108.L100401, Kim2024}. Our regression model, illustrated in Fig.~\ref{fig1}(a), accurately estimates the parameters of the magnetic Hamiltonian from magnetic domain images generated through atomistic spin simulations. This model resolves an inverse problem associated with inferring the parameters from the observed data, a feat beyond the reach of established theoretical methods \cite{Hejazi10721, PhysRevB.104.L100406, PhysRevB.108.174440, Tong2018, Yang2023, PhysRevB.103.L140406, PhysRevResearch.3.013027, Akram2021, PhysRevB.104.014410, Ghader2022, Fumega_2023, Zheng2022, Kim2023, PhysRevB.108.L100401, Kim2024}. Our “generative model,” depicted in Fig.~\ref{fig1}(b), excels at producing magnetic domain images based on predefined magnetic parameters. This model, designed with deterministic characteristics, differs from typical generative frameworks by consistently producing a unique ground-state structure from given parameters. It effectively handles the complexity of the magnetic Hamiltonian without compromising accuracy, thereby reducing the need for resource-intensive atomistic spin simulations \cite{Zheng2022, Kim2023, Yang2023, PhysRevB.108.L100401, Kim2024, Akram2024}. These advancements not only extend the applicability of DNN techniques to twisted van der Waals magnets but also streamline future investigations into these captivating systems.

\section{Methods}

\subsection{Dataset generation}

\begin{figure}[t!]
    \centering
    \includegraphics[width=.9\textwidth]{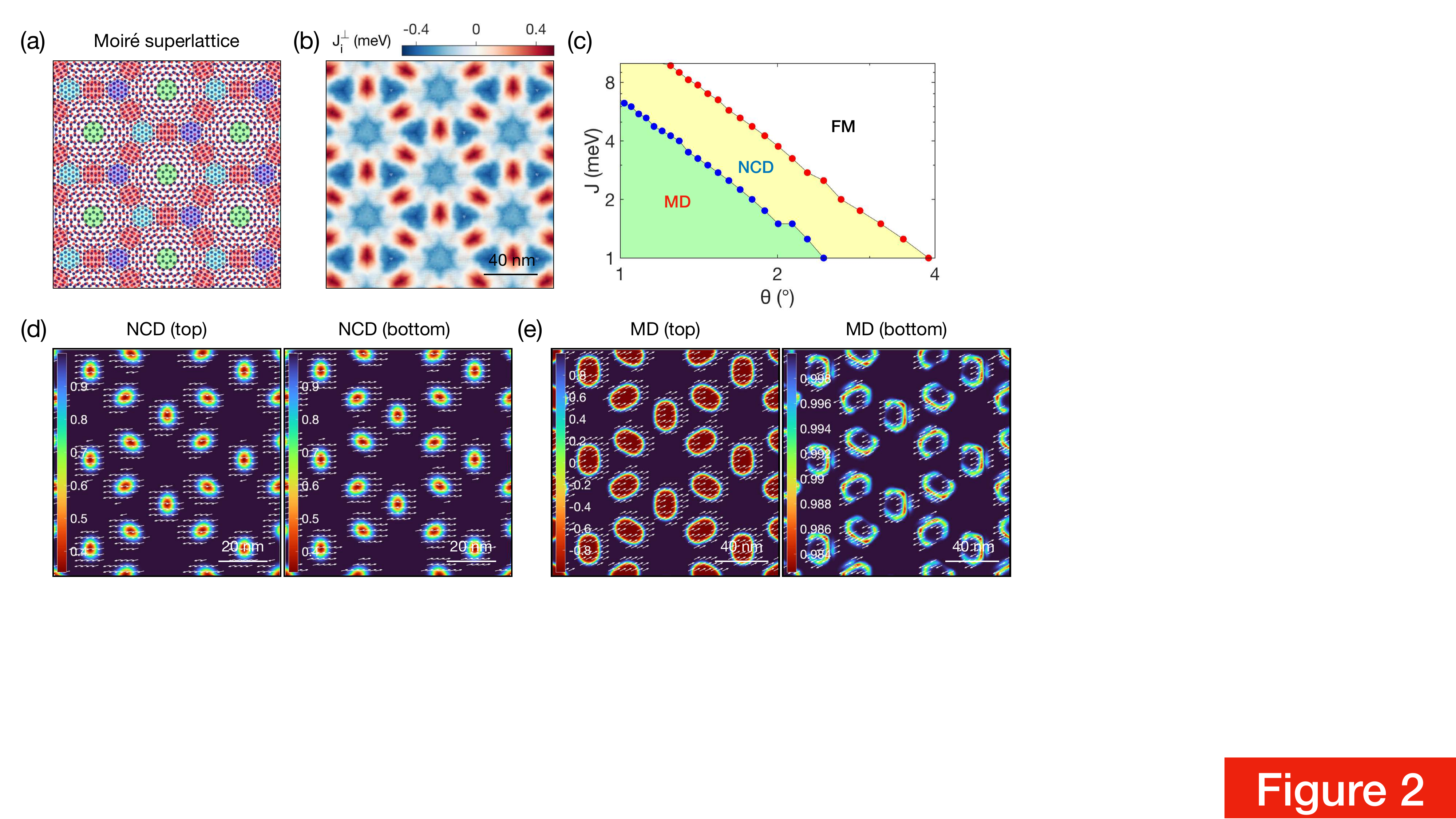}
    \caption[Magnetic domain array]{Twisted bilayer CrI\textsubscript{3} and its magnetic domain images employed in this study. (a) Moiré superlattice for a twist angle $\theta = 5.08$\textdegree{}. The colored circles denote local stacking patterns, including AA (green), AB (blue), BA (cyan), and monoclinic (red) \cite{Kim2023}. (b) Modulation of the local interlayer exchange energy ($J_i^\perp=\sum_jJ_{ij}^\perp$) calculated for $\theta=1.02$\textdegree{}. Red and blue colors indicate antiferromagnetic and ferromagnetic (FM) interlayer exchange couplings, respectively. (c) Zero temperature magnetic phase diagram for $A=0.2$ meV as a function of $\theta$ and intralayer exchange ($J$), showing the FM phase (white), noncollinear domain (NCD) phase (yellow), and magnetic domain (MD) phase (green). (d–e) Magnetic configurations illustrating nanoscale magnetic domain arrays in the NCD (d) and MD phases (e), respectively. Left and right panels in each plot correspond to the top and bottom layers, respectively. The color scale depicts the out-of-plane magnetization directions, and the arrows indicates the in-plane magnetization directions. In (d), $J=2$ meV, $A=0.2$ meV and $\theta=1.89$\textdegree{} are utilized. In (e), $J=2$ meV, $A=0.2$ meV and $\theta=1.02$\textdegree{} are utilized.}
    \label{fig2}
\end{figure}

To generate magnetic domain images, we employ twisted bilayer CrI\textsubscript{3} \cite{Hejazi10721, PhysRevResearch.3.013027, Akram2021, Ghader2022, Zheng2022, Fumega_2023, Kim2023, Yang2023, PhysRevB.108.174440} in commensurate moiré superlattice structures (Fig.~\ref{fig2}(a)). The magnetic behavior of this system is effectively described using a Heisenberg spin model \cite{PhysRevResearch.3.013027, Zheng2022, Fumega_2023, Kim2023}:
\begin{align} \label{eq:spinH}
    H = & -J\sum_{l=t,b} \sum_{\langle i, j\rangle} \bm{S}_{i}^l \cdot \bm{S}_{j}^l - A \sum_{l=t,b} \sum_{i} \big(\bm{S}_{i}^l\cdot\hat{z}\big)^2 + \sum_{i, j} J_{ij}^\perp \bm{S}_{i}^t \cdot \bm{S}_{j}^b.
\end{align}
Here, $\bm{S}_{i}^{l}$ represents the spin at site $i$ on the top layer ($l=t$) and the bottom layer ($l=b$). $J>0$ represents intralayer ferromagnetic (FM) exchange interactions. $A>0$ represents single-ion anisotropy favoring out-of-plane magnetization. $J_{ij}^\perp$ represents interlayer exchange interactions derived from ab-initio calculations on bilayer CrI\textsubscript{3} \cite{Kim2023}. The coupling type of $J_{ij}^\perp$ varies between antiferromagnetic and FM depending on the local stacking pattern in the moiré superlattice (Fig.~\ref{fig2}(b)).

Figure~\ref{fig2}(c) illustrates three magnetic phases inherent in this spin model \cite{Zheng2022, Kim2023}: (i) the FM phase, (ii) the noncollinear domain (NCD) phase, and (iii) the magnetic domain (MD) phase. The FM phase exhibits a perpendicular, uniform spin configuration. In the NCD phase, magnetic domains form in both layers, with spins tilted horizontally in opposite directions (Fig.~\ref{fig2}(d)). On the other hand, in the MD phase, domains form in one layer with opposite out-of-plane polarization compared to the background FM region (Fig.~\ref{fig2}(e)). In the other layer, the magnetization largely maintains an FM order, but slight spin tilting occurs around localized domain wall regions. Both NCD and MD phases exhibit moiré periodicity, leading to the periodic domain patterns (Fig.~\ref{fig2}(d–e)), which we refer to as “nanoscale magnetic domain arrays” \cite{Kim2024}.

We generate magnetic domain images through atomistic spin simulations \cite{Kim2023} using Eq.~\eqref{eq:spinH}. These simulations incorporate randomly generated parameter sets ($\theta$, $J$, $A$) spanning from 1.01\textdegree{} to 3.89\textdegree{} for $\theta$, 1 to 10 meV for $J$, and 0.01 to 0.3 meV for $A$, encompassing $(J, A) \approx (2.0~\textrm{meV}, 0.2~\textrm{meV})$ pertinent to CrI\textsubscript{3} \cite{PhysRevX.8.041028}. The values of $J_{ij}^\perp$ are fixed, determining the overall energy scale of the magnetic parameters. For each parameter set, we obtain a magnetic ground state and generate two images for each state, one for the top and the other for the bottom layer. These images depict out-of-plane magnetization within a fixed length dimension of 100 nm. Additionally, we consistently omit the FM phase from the dataset to prevent potential overfitting issues \cite{Hawkins2004}, albeit resulting in biased parameter distributions. Consequently, our dataset comprises 17,292 paired images, capturing morphological modifications of nanoscale magnetic domain arrays within the NCD and MD phases corresponding to the parameter variations. Further details about the dataset generation process can be found in Appendix~\ref{app:dataset}.

\subsection{Deep learning algorithms}

\subsubsection{Regression model}

We implement our regression model using a customized, fully connected network. Three networks are employed, each predicting one assigned parameter, collectively yielding the entire parameter set ($\theta$, $J$, $A$). Each network consists of four hidden layers, employing the Gaussian error linear unit (GELU) activation function \cite{gelu}. The input layer of the network receives two images, each sized at $(200, 200)$, illustrating magnetic domains at the top and bottom layers. These inputs separately undergo dimensional reduction to $512$ features within the first two hidden layers, which are then concatenated within a single list with a length of $1024$. This list is subsequently processed through the two remaining hidden layers. The final output layer yields a predicted parameter using a linear activation function. For a visual presentation of the network architecture, refer to Appendix~\ref{app:network_structures}.

We train the networks to minimize the discrepancy between predicted and actual Hamiltonian parameters, employing mean squared error (MSE) as a loss function. The Adam optimizer is employed with a dynamic learning rate ranging from $10^{-4}$ to $10^{-3}$, and a batch size of $128$ was chosen to facilitate efficient gradient updates. Additionally, we incorporate early stopping as a regularization technique to prevent overfitting and enhance generalization performance during training \cite{Hawkins2004}. These carefully considered choices in training parameters are made to ensure optimal performance. Two randomly partitioned sets (13,836 samples and 3,459 samples) from the full dataset are utilized as training and test sets.

\subsubsection{Generative model}

Our generative model is implemented using a customized, fully connected network. The network consists of two hidden layers, employing the GELU activation function. In the input layer, the network receives three parameters ($\theta$, $J$, $A$), which are processed through two hidden layers, and culminated in a list of 77,618 features in the output layer. This list is subsequently reshaped into two pixel arrays with dimensions (197, 197), representing magnetic domain images that delineate the out-of-plane components of spins in both the top and bottom layers. For a detailed visual representation of the network architecture, please refer to Appendix~\ref{app:network_structures}.

We train the network to minimize the discrepancy in pixel values between true and predicted magnetic domain images, utilizing the mean absolute error (MAE) as our chosen loss function. The detailed training process aligns with the regression model, involving the Adam optimizer with a dynamic learning rate ranging from $10^{-4}$ to $10^{-3}$, a batch size of $128$, and the implementation of early stopping as a regularization technique. The generative model is trained and tested using the same training and test datasets that are used in the regression model.

\subsection{Statistical error analysis} \label{sec:error_analysis}

Our statistical error analysis in both regression and generative models involves normalizing $\theta$, $J$, and $A$ as follows:
\begin{align}
    \theta_N=&\;\frac{\theta-\textrm{min}(\theta)}{\textrm{max}(\theta)-\textrm{min}(\theta)}, \nonumber \\
    J_N=&\;\frac{J-\textrm{min}(J)}{\textrm{max}(J)-\textrm{min}(J)}, \nonumber \\
    A_N=&\;\frac{A-\textrm{min}(A)}{\textrm{max}(A)-\textrm{min}(A)}. \label{eq:norm_params}
\end{align}
This normalization ensures that all parameters fall within the $[0,1]$ range, facilitating consistent treatment across different parameters \cite{doi:10.1126/sciadv.abb0872}. Furthermore, the Euclidean distance from each data point to the phase boundary is calculated using the following expression:
\begin{align}
    R=\sqrt{(\theta_N-\theta_N^*)^2+(J_N-J_N^*)^2+(A_N-A_N^*)^2}.
\end{align} 
Here, $(\theta_N,J_N,A_N)$ and $(\theta_N^*,J_N^*,A_N^*)$ represent a data point in the test set and its corresponding nearest point on the FM–NCD or NCD–MD phase boundary.

In the regression model, the MAE values for each normalized parameter are calculated as:
\begin{align}
    \Delta\theta_N=&\;\left|\theta_N^\textrm{Pred.}-\theta_N^\textrm{True}\right|, \nonumber\\
    \Delta J_N=&\;\left|J_N^\textrm{Pred.}-J_N^\textrm{True}\right|, \nonumber\\
    \Delta A_N=&\;\left|A_N^\textrm{Pred.}-A_N^\textrm{True}\right|, \label{eq:norm_params_MAE}
\end{align} 
where ‘True’ and ‘Pred.’ indicate the true and predicted values in the test set, respectively. Furthermore, our principal component analysis \cite{Jolliffe:1986, doi:10.1126/sciadv.abb0872} incorporates error vectors, defined as:
\begin{align}
    X = (\Delta \theta_N, \Delta J_N, \Delta A_N).
\end{align}
The covariance matrix of the error vectors is computed using the following expression:
\begin{align}
    K_{ij} = E[(X_i-E[X_i])(X_j-E[X_j])].
\end{align}
Here, $X_{i,j}$ for $i,j=1,2,3$ represents each component in the error vector $X$. $E[X_i]$ indicates the average value of $X_i$ among different samples in the test set. The eigenvectors and eigenvalues of $K_{ij}$ are computed, and the principal components are selected based on the magnitude of their eigenvalues. Variance ratios are computed for each component using $ \frac{\lambda_i}{\sum_{i=1}^3\lambda_i}\times 100 $, where $\lambda_i$ denotes the eigenvalue for each component.

\subsection{Generation of noisy test sets} \label{sec:noisy test sets}

We generate noisy test sets from the original test set by introducing Gaussian-white noises according to the equation:
\begin{align}
     z_i = z_i^0 + \delta z_i. \label{eq:pixel_value}
\end{align}
Here, $z_i^0$ and $z_i$ represent the pixel values at the $i$-th pixel in the original and noise-injected images, respectively. The injected noise $\delta z_i$ follows a Gaussian distribution:
\begin{align}
     P(\delta z_i) = \frac{1}{\sigma\sqrt{2\pi}}\exp{\bigg(-\frac{\delta z_i^2}{2\sigma^2} \bigg)}. \label{eq:noise_dist}
\end{align}
The variance $\sigma^2$ of the distribution is effectively controlled by adjusting a signal-to-noise ratio (SNR), defined as $R_\textrm{SNR}= P_\textrm{signal} / P_\textrm{noise}$. Here, $P_\textrm{signal}=\frac{1}{N_\textrm{pixel}}\sum_{i=1}^{N_\textrm{pixel}}\big|z_i^0\big|^2$ and $P_\textrm{noise}=\frac{1}{N_\textrm{pixel}}\sum_{i=1}^{N_\textrm{pixel}}\big|\delta z_i\big|^2$ represent signal and noise power, respectively. The adjustment using $R_\textrm{SNR}$ ensures consistency across different images with various $P_\textrm{signal}$ levels \cite{doi:10.1126/sciadv.abb0872}. Given $P_\textrm{signal}$ obtained from a noiseless image and a specified value of $R_\textrm{SNR}$, we determine $\sigma^2$ through the relationship:
\begin{align}
    \sigma^2 = P_\textrm{signal} / R_\textrm{SNR},
\end{align}
which is derived from $\sigma^2 \approx P_\textrm{noise}$, valid for a sufficiently large $N_\textrm{pixel}$, the number of pixels in the image. In the generation process of noisy datasets, the SNR is systematically varied over a wide range of $R_\textrm{SNR}$ using a decibel scale:
\begin{align}
    R_\textrm{SNR}^\textrm{dB} = 10\log_{10}(R_\textrm{SNR}), \label{eq:R_dB}
\end{align}
where $R_\textrm{SNR}^\textrm{dB}$ is tuned from $-5$ dB to $60$ dB.

\section{Results}

\subsection{Regression model: parameter estimation from domain images}

\subsubsection{Validation of trained networks}

\begin{figure}[t!]
    \centering
    \includegraphics[width=.75\textwidth]{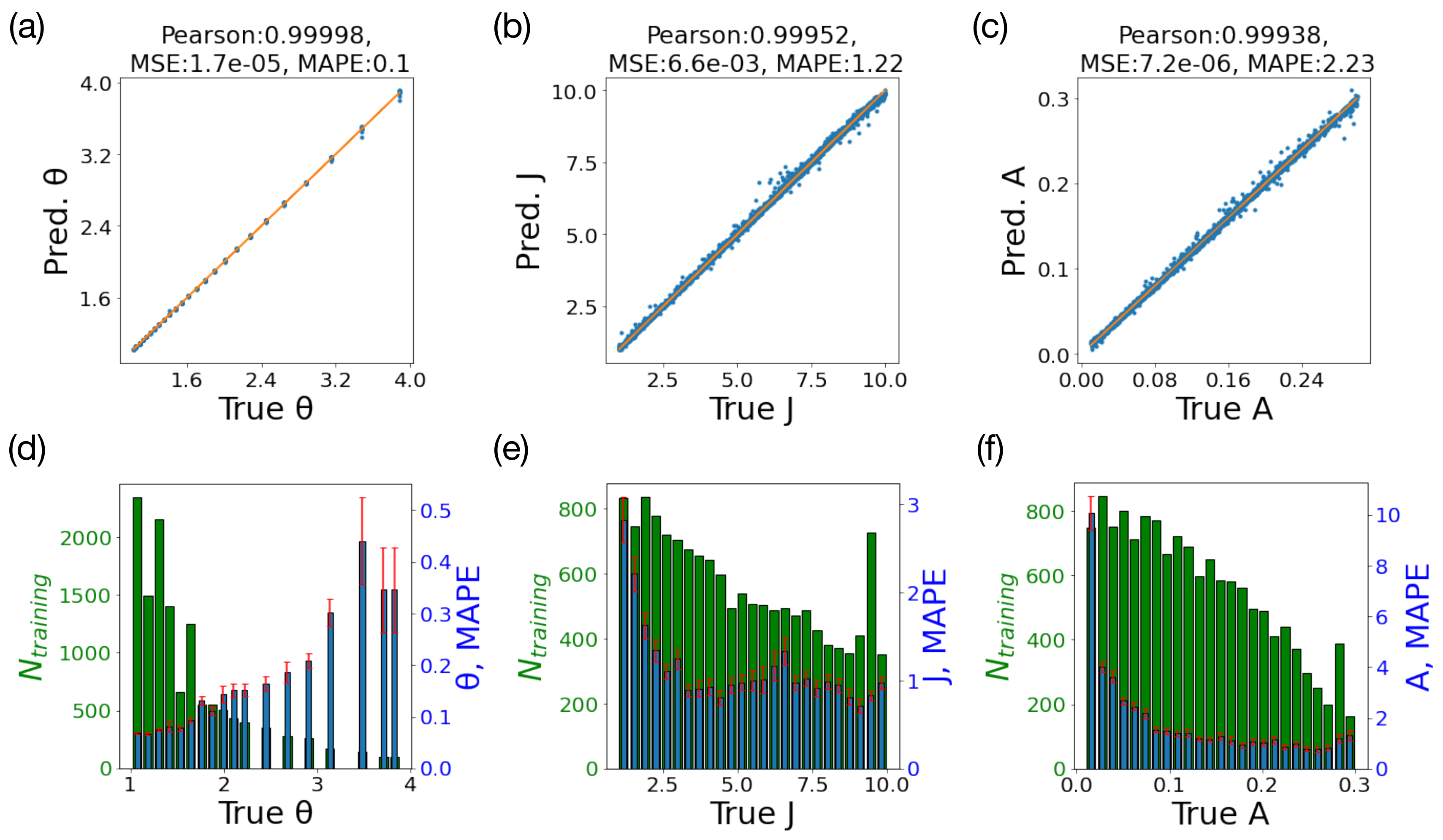}
    \caption[Regression model]{Performance of the trained networks in the regression model. (a–c) Parameter estimation results using the test set. The $x$-axis denotes the true values used in the simulation, while the $y$-axis represents the model's estimated values. The yellow lines ($y=x$) represent the ideal predictions. Evaluation metrics include Pearson correlation coefficients (labeled as Pearson), mean squared error (MSE), and mean absolute percentage error (MAPE). (d–f) Profiles of the parameter estimation errors derived from (a–c). Green bars depict the count of samples ($N_\textrm{training}$) within each parameter range in the training set. Blue bars represent the MAPE values for the estimated parameters. Error bars signify the standard errors of the MAPE values. The units of degree, meV, and meV are utilized for $\theta$, $J$, and $A$, respectively. }
    \label{fig3}
\end{figure}

Figure~\ref{fig3} presents the parameter estimation results from our trained networks on the test set. The $\theta$ estimation demonstrates precise agreement between predicted and true values, evidenced by high-level scores in evaluation metrics (Fig.~\ref{fig3}(a)). Further analysis reveals consistently low mean absolute percentage errors (MAPEs) across the entire $\theta$ range (Fig.~\ref{fig3}(d)). Particularly noteworthy is the observation that MAPE values remain below 0.1 in the small twist angle regime ($\theta \lesssim 1.5 $\textdegree{}). This capability indicates the network's proficiency in estimating $\theta$, underscoring its promising applications to small twist angle systems, which are frequently involved in experimental investigations \cite{Song2021, Xie2022, Xu2022, Xie2023, Cheng2023}.

Estimations for $J$ and $A$ also exhibit consistent linear relationships between predicted and true values, showing small MAPE values of 1.22 and 2.23 for $J$ and $A$, respectively (Fig.~\ref{fig3}(b–c)). Further examination reveals that the MAPE value for $J$ and $A$ tend to increase as their true values decrease, while high accuracy is maintained for sufficiently large $J$ and $A$ values (Fig.~\ref{fig3}(e–f)). Specifically, in the vicinity of $(J, A) \approx (2.0~\textrm{meV}, 0.2~\textrm{meV})$ pertinent to CrI\textsubscript{3} \cite{PhysRevX.8.041028}, the MAPE values stand at small numbers, 1.9 and 0.7 for $J$ and $A$, respectively. These results showcase the networks' reliable predictive capabilities, specifically in magnetic systems with moderate values of $J$ and $A$, including CrI\textsubscript{3}.

\subsubsection{Statistical error analysis: error correlation between $J$ and $A$} \label{sec:error_analysis_regression}

\begin{figure}[t!]
    \centering
    \includegraphics[width=.8\textwidth]{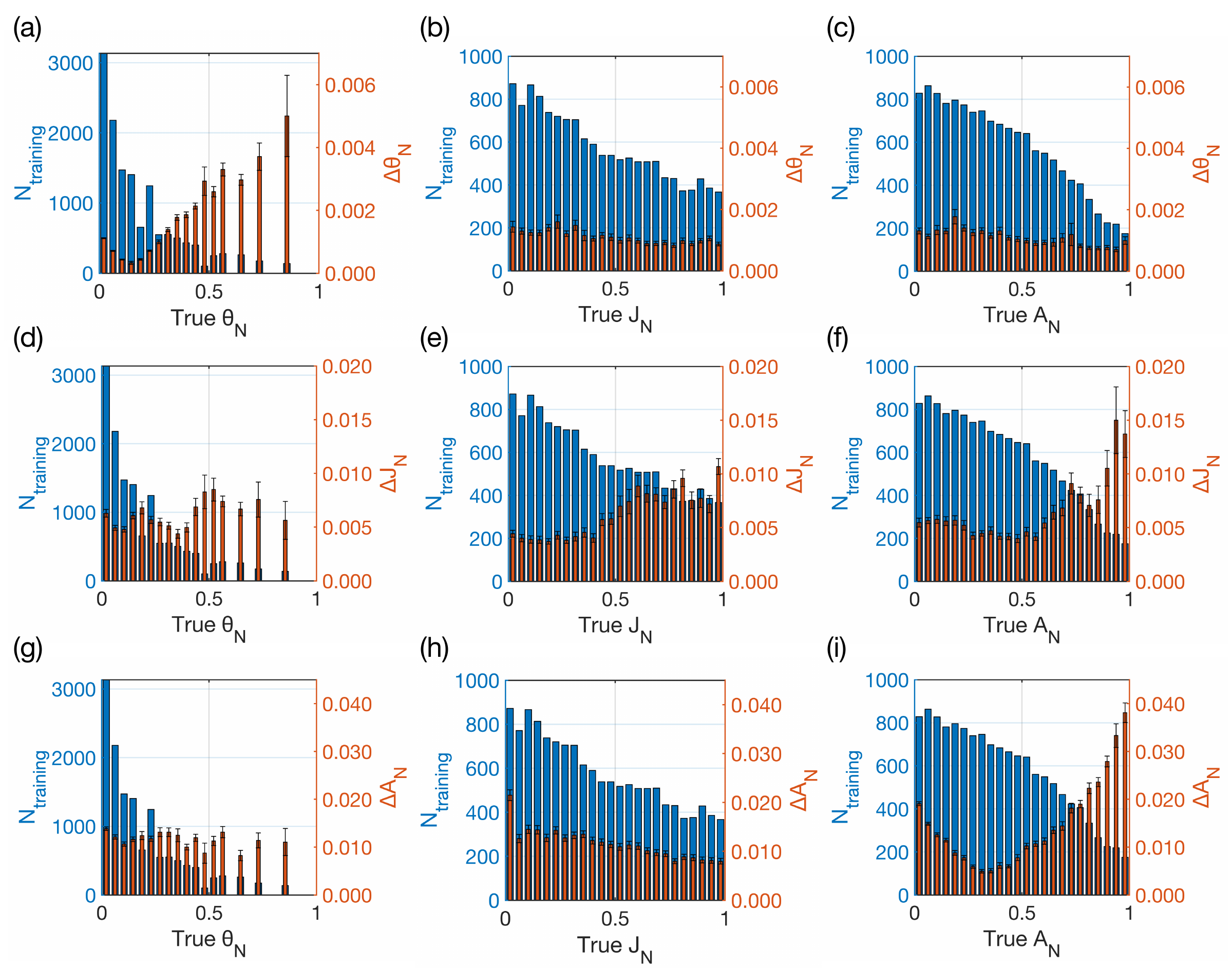}
    \caption[Error in regression 1]{Profiles of parameter-estimation errors concerning parameter variations. The $x$-axis denotes the normalized parameters ($\theta_N$, $J_N$, and $A_N$). Blue bars represent the sample count ($N_\textrm{training}$) in the training set. Red bars depict the mean absolute error (MAE) values ($\Delta \theta_N$, $\Delta J_N$, and $\Delta A_N$). Error bars indicate the standard errors of the MAE values.}
    \label{fig4}
\end{figure}

To delve deeper into the errors in parameter estimation, we conduct a statistical error analysis on these errors. Figure~\ref{fig4} illustrates the profiles of the MAE values ($\Delta \theta_N$, $\Delta J_N$, $\Delta A_N$) and the sample count ($N_\textrm{training}$) in the training set, concerning the variation of the normalized parameters ($\theta_N$, $J_N$, $A_N$). Our finding reveals that the MAE values tend to increase as their corresponding values of $N_\textrm{training}$ decrease (Fig.~\ref{fig4}(a,e,i)). This tendency demonstrates that sample scarcity is one of the primary factors contributing to these errors, aligning with common expectations in machine learning modeling. In addition to this, we notice an unusual rise in $\Delta A_N$ within a range characterized by a small $A_N$ value, despite an increase in $N_\textrm{training}$ (Fig.~\ref{fig4}(i)). One of the possible explanations for this is that the rapid alterations in magnetic domain structures within this range \cite{Kim2023} make it challenging for the networks to learn these patterns from the limited dataset.

We observe no discernible correlations between $\Delta \theta_N$–$J_N$, $\Delta \theta_N$–$A_N$, $\Delta J_N$–$\theta_N$, and $\Delta A_N$–$\theta_N$ (Fig.~\ref{fig4}(b,c,d,g)), suggesting that variations in $J_N$ and $A_N$ do not significantly affect the estimation of $\theta_N$, and vice versa. In contrast, noticeable correlations exist between $\Delta J_N$–$A_N$ and $\Delta A_N$–$J_N$ (Fig.~\ref{fig4}(f,h)), indicating mutual impacts on their parameter estimation. These contrasting tendencies may stem from the distinct roles among different parameters in determining nanoscale magnetic domain arrays; while $\theta_N$ solely dictates the arrays' periodicity, $J_N$ and $A_N$ collectively influence the detailed morphology of individual domains \cite{Kim2023}. Consequently, these results suggest that the networks recognize these two factors—the periodicity and domain shape of the domain arrays—and process them independently, despite not being explicitly incorporated into the algorithm. In other words, our model effectively exploits the networks' capability to learn from data without explicit manual programming, which is one of the primary benefits of deep learning techniques.

\begin{figure}
    \includegraphics[width=0.36\textwidth]{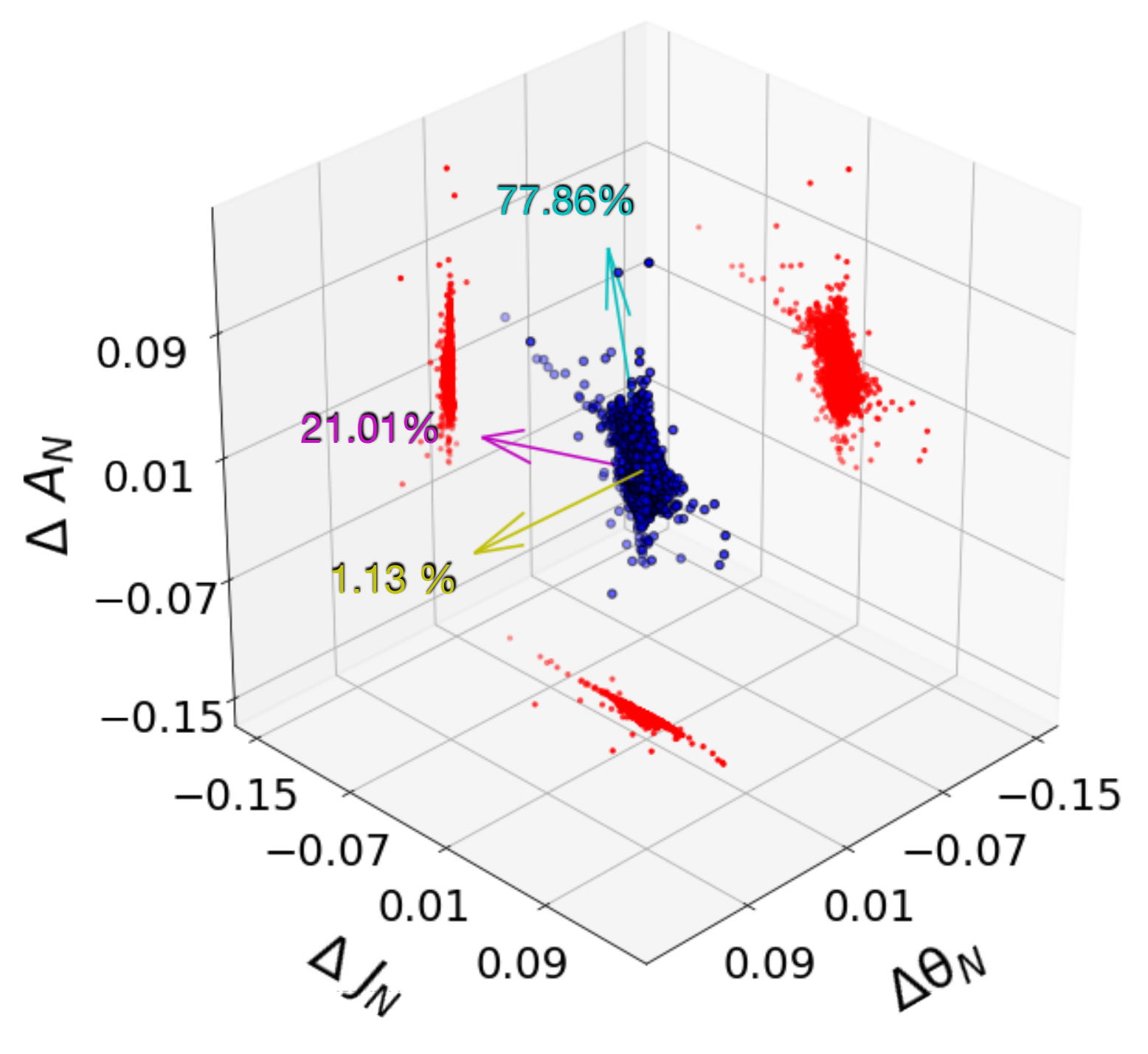}
    \caption[PCA]{Visualization of error vectors and their principal components. Blue dots denote the error vectors ($\Delta \theta_N$, $\Delta J_N$, $\Delta A_N$), while red dots indicate their projection onto the respective planes. Cyan, magenta, and yellow arrows denote the first, second, and third principal components, given by $(-0.017,-0.227,0.974)$, $(0.019,-0.974,-0.226)$, and $(0.974,-0.226,0.021)$, respectively. The numbers next to each arrow specify the variance ratio of the corresponding component.} \label{fig5}
\end{figure}

To gain deeper insights into the correlation among errors, we conduct a principal component analysis on the error vectors ($\Delta \theta_N, \Delta J_N, \Delta A_N$), as depicted in Fig.~\ref{fig5}. Notably, the error distribution appears highly concentrated within a two-dimensional subspace delineated by $\Delta J_N$ and $\Delta A_N$, reflecting the relatively large values of $\Delta J_N$ and $\Delta A_N$ compared to $\Delta \theta_N$ (Fig.~\ref{fig3}(d–f)). Moreover, within this subspace, the distribution of $\Delta J_N$ and $\Delta A_N$ manifests notable asymmetry, as demonstrated by the principal component $(-0.017,-0.227,0.974)$, which accounts for 77.86\% of the observed variance. This predominant axis, intertwined with $\Delta J_N$ and $\Delta A_N$, signals a significant relationship between the two errors \cite{doi:10.1126/sciadv.abb0872}, succinctly expressed as:
\begin{align}
    \Delta J_N \approx -0.233 \Delta A_N.
\end{align}
This relationship suggests a tendency for underestimation in $\Delta J_N$ to coincide with overestimation in $\Delta A_N$, and vice versa. Moreover, this correlation may elucidate the observed error magnification in $\Delta J_N$ and $\Delta A_N$ (Fig.~\ref{fig3}(e–f)), where errors in one variable propagate and exacerbate those in another, resulting in more pronounced deviations \cite{10.5555/2621980}.

The observed correlation between $\Delta J_N$ and $\Delta A_N$ may offer valuable physical insights into the system. Specifically, it suggests an intricate interplay between $J$ and $A$ in shaping the detailed morphology of nanoscale magnetic domain arrays. One conceivable scenario is that $J$ and $A$ operate interchangeably, \textit{i.e.,} an increase in one parameter compensates for a decrease in the other, thereby preserving the overall structure. This possibility aligns with previous findings in the interaction between magnetic dipolar interactions and single-ion anisotropy \cite{doi:10.1126/sciadv.abb0872}. While elucidating the precise mechanism lies beyond the scope of this study, the specific relationship uncovered here may provide valuable insights into this pursuit.

\begin{figure}[t!]
    \centering
    \includegraphics[width=.8\textwidth]{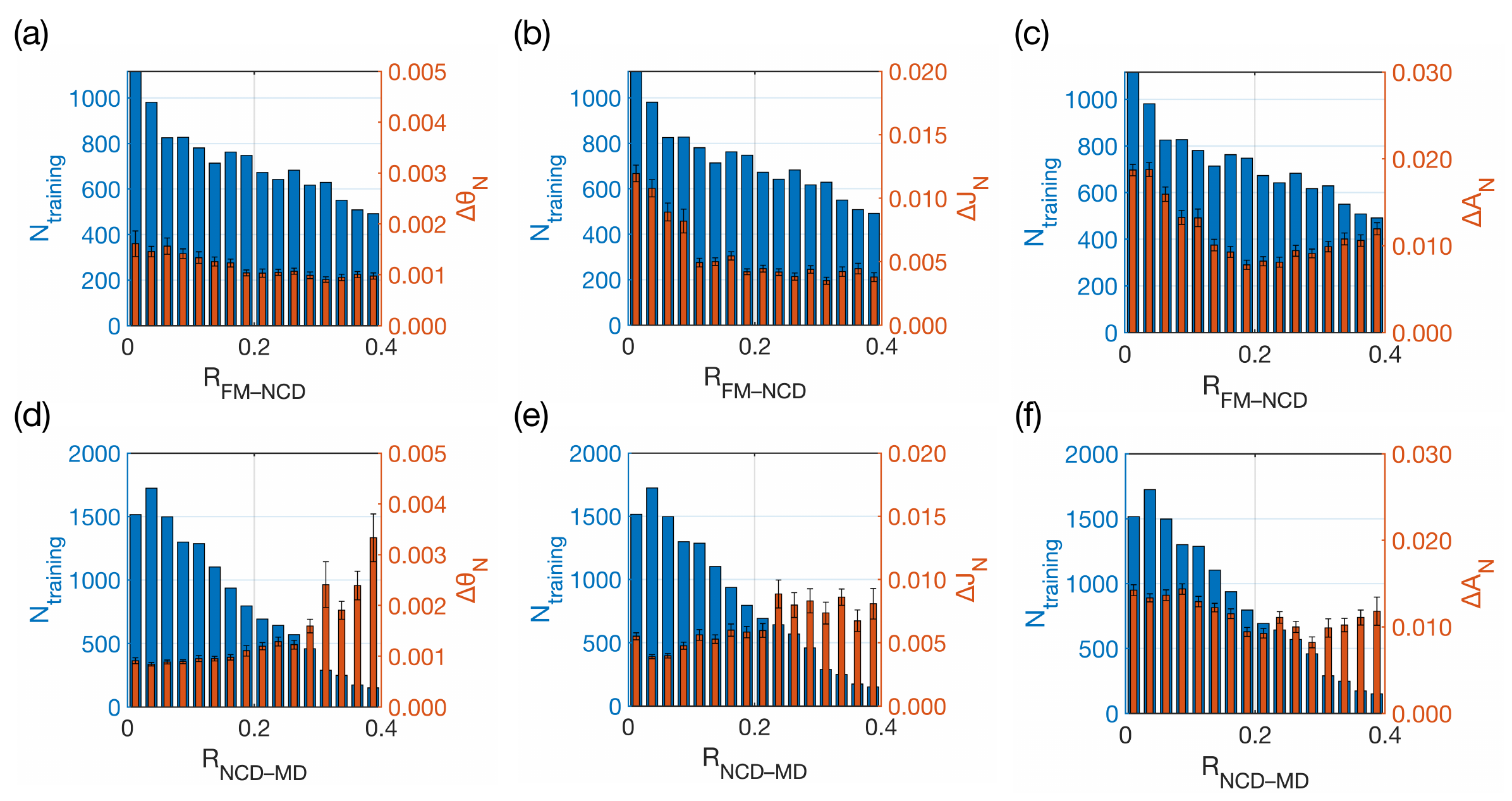}
    \caption[Error in regression 2]{Profiles of parameter-estimation errors near phase boundaries. The $x$-axis denotes the Euclidean distance from each data point to the FM–NCD or NCD–MD phase boundary ($R_\textrm{FM–NCD}$ or $R_\textrm{NCD–MD}$). Blue bars depict the sample count ($N_\textrm{training}$) in the training set. Red bars depict the MAE values ($\Delta \theta_N$, $\Delta J_N$, and $\Delta A_N$). Error bars signify the standard errors of the MAE values.}
    \label{fig6}
\end{figure}

For a thorough understanding of the errors, we evaluate the network's performance near phase boundaries, where sharp variations in magnetic domain structures can significantly affect parameter estimation. Figure~\ref{fig6} illustrates the profiles of MAE values concerning the Euclidean distance from each data point to the FM–NCD or NCD–MD phase boundaries that quantify the degree of separation from these phase boundaries. We observe a notable increase in MAE values as the Euclidean distance to the FM–NCD phase boundary decreases (Fig.~\ref{fig6}(a–c)), indicating amplified errors near this boundary. This trend contrasts with the growing number of samples within this region, suggesting that the errors are more influenced by the physical aspects of this transition rather than sample scarcity. However, no discernible correlation is found between the MAE values and the Euclidean distance to the NCD–MD phase boundary (Fig.~\ref{fig6}(d–f)). Consequently, we infer that the FM–NCD transition plays a pivotal role in magnifying errors in parameter estimation.

From these results, we deduce that errors in parameter estimation within the regression model may stem from three potential sources of magnification: (i) limited sample availability in the training set, (ii) error correlation, and (iii) the FM–NCD transition. Although the hierarchy among these factors and their potential interplay lie beyond the scope of this study, these identifications offer valuable insights for implementing meticulous treatments aimed at enhancing predictive accuracy.

\subsubsection{Robustness against noise injections}

\begin{figure}[t]
    \centering
   \includegraphics[width=0.85\textwidth]{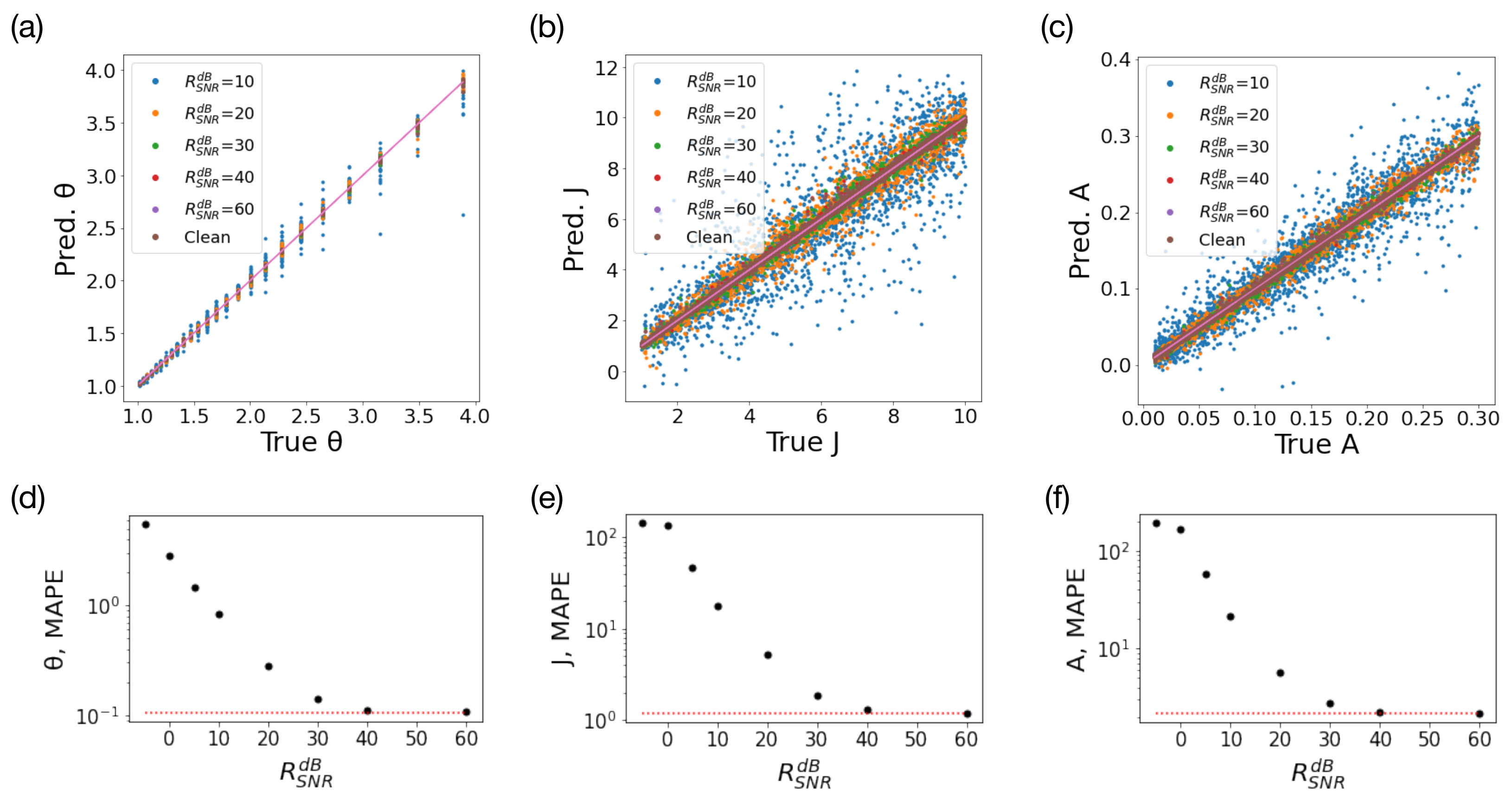}
    \caption{Performance of the trained networks in the regression model in the presence of noises. (a–c) Parameter estimation results derived from noisy test sets in comparison to the noiseless original test set (denoted by Clean). Here, $R_\textrm{SNR}^\textrm{dB}$ denotes the signal-to-noise ratio, effectively reflecting the noise level on a decibel scale. The $x$-axis denotes the true values used in the simulation, while the $y$-axis represents the model's estimated values. The pink lines ($y=x$, denoted by True) represent the ideal predictions. (d–f) MAPE values for the estimated parameters in different $R_\textrm{SNR}^\textrm{dB}$ levels. The markers denote MAPE values for each parameter. The red dashed line depicts the MAPE values for the Clean test set.}
    \label{fig7}
\end{figure}

For real-world applications, including noisy experimental data, we evaluate the regression model's performance under noisy conditions. Figure~\ref{fig7} presents the parameter estimation results on noisy test sets obtained from the networks trained on the noiseless training set. The $\theta$ estimation sustains a consistent linear relationship between predicted and true values across the entire $\theta$ range (Fig.~\ref{fig7}(a)). Moreover, the MAPE value remains steadfastly low at 5\% to the highest noise level of $R_\textrm{SNR}^\textrm{dB}=-5$ dB (Fig.~\ref{fig7}(d)). Consequently, this result supports that the network retains its proficiency in accurately extracting $\theta$ from magnetic domain images containing substantial noises.

Estimation for $J$ and $A$ also display consistent linear relationships between predicted and true values (Fig.~\ref{fig7}(b–c)), although their precision considerably degrades as the noise level increases. Specifically, within a weak noise regime ($R_\textrm{SNR}^\textrm{dB} \geq 20$ dB), the MAPE values are sustained at relatively small values less than $4\%$ and $6\%$ for $J$ and $A$, respectively (Fig.~\ref{fig7}(e–f)). However, upon entering a strong noise regime ($R_\textrm{SNR}^\textrm{dB} < 20$ dB), the MAPE values steeply increase, reaching considerable levels exceeding $20\%$ for both $J$ and $A$ at $R_\textrm{SNR}^\textrm{dB}=10$ dB. Consequently, the networks' capability to make meaningful predictions in $J$ and $A$ is practically restricted to $R_\textrm{SNR}^\textrm{dB}=20$ dB.

We further validate the regression model under thermal fluctuations by employing noisy test sets generated through the solution of the stochastic Landau–Lifshitz–Gilbert equation \cite{Aron_2014}. Our analysis reveals that the trained network maintains its accurate predictive capabilities for estimating $\theta$ and $J$ up to the highest investigated temperature of 5 K. However, estimating $A$ from the noisy test sets has proven highly challenging due to the dominant energy scale of the random thermal field compared to $A$. Consequently, the intricate morphological features associated with $A$ could be overshadowed by the overwhelming random thermal field, leading to the network struggling to accurately estimate $A$ from these features.

These findings suggest that despite being trained using a noiseless training set, the network could be applied to data subject to thermal noises, such as experimental data, at least for estimating $\theta$ and $J$. It's worth mentioning that one can expect higher performance in parameter estimation in the MD phase due to the robustness of magnetic structures against thermal noises within this phase. For a more detailed analysis, please refer to Appendix~\ref{app:thermal_noises}.

\subsection{Generative model: domain image generation from parameters}

\subsubsection{Validation of trained networks}

\begin{figure}[t!]
    \centering
    \includegraphics[width=.85\textwidth]{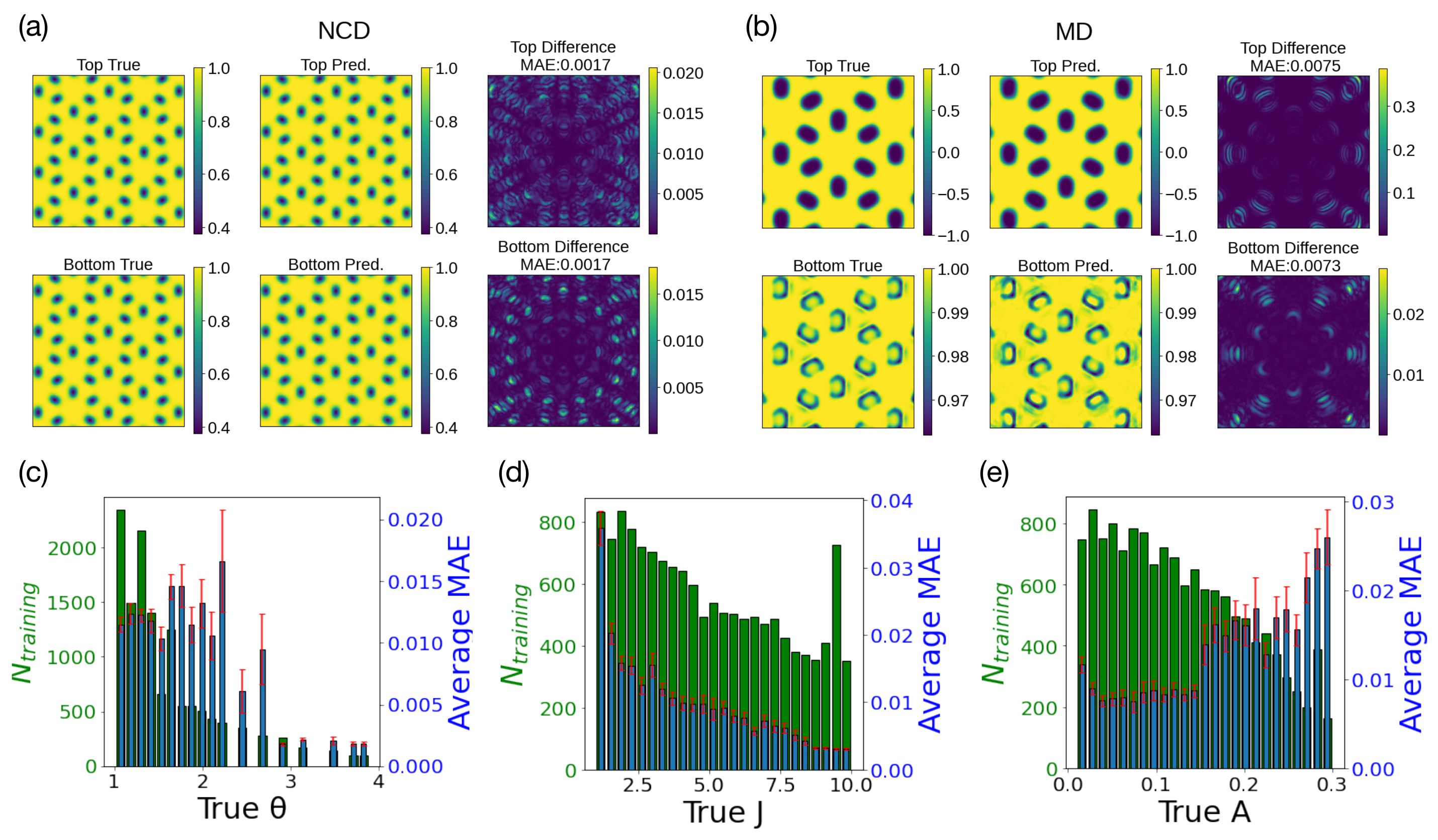}    
    \caption[Generative model]{Performance of the trained network in the generative model. (a–b) Examples of predicted magnetic domain images corresponding to (a) the NCD and (b) MD phases. The color scale indicates the magnitude and direction of out-of-plane magnetization. The first and second columns display the true and predicted images, and the third column shows the MAE values of the out-of-plane magnetization between the two images. The first and second rows represent the top and bottom layers, respectively. The parameter sets $(\theta, J, A)=$ (3.15, 5.36, 0.14) and $(\theta, J, A)=$ (1.16, 6.00, 0.04) are utilized for (a) and (b), respectively. (c–e) Profiles of the average MAE values obtained from the predicted images in the test set. Green bars depict the count of samples ($N_\textrm{training}$) within each parameter range in the training set. Blue bars represent the average MAE values of the predicted images. Error bars signify the standard errors of the average MAE values. The degree, meV, and meV units are utilized for $\theta$, $J$, and $A$, respectively.}
    \label{fig8}
\end{figure}

Figure~\ref{fig8} illustrates the magnetic domain image prediction results from our trained network on the test set. The predicted images precisely replicate the intricate pattern of the nanoscale magnetic domain arrays in the true images, capturing details such as the domains' location, shape, and size (the first and second columns in Fig.~\ref{fig8}(a–b)). Additionally, the magnitude of local magnetization exhibits precise agreement between the two images over the majority of the region (the third column in Fig.~\ref{fig8}(a–b)), as evidenced by small average MAE values of approximately $\sim0.002$ and $\sim0.007$ for the NCD and MD phases, respectively. This precise agreement persists consistently across the entire parameter ranges, with average MAE values remaining below $0.03$ (Fig.~\ref{fig8}(c–e)). These observations affirm the proficient performance of the trained network in accurately predicting magnetic domain images from given parameters.

Despite its overall decent performance, the network shows a notable discrepancy in predicting magnetic domain images between the NCD and MD phases. In the NCD phase, the MAE value between true and predicted images remains consistently low, less than $0.02$ (the third column in Fig.~\ref{fig8}(a)). Conversely, in the MD phase, relatively large MAE values exceeding $0.3$ appear in a layer where magnetic domains form (the third column in Fig.~\ref{fig8}(b)). A closer examination reveals a concentration of these high MAE values within the local domain-wall regions, where sharp alterations in magnetization occur. This is in contrast to the NCD phase, which lacks domain walls and exhibits relatively smooth variations across the entire region. This observation leads us to posit that the pronounced discrepancy in the MD phase arises from the network's challenge in learning sharp variations within domain wall regions from the limited dataset.

\subsubsection{Statistical error analysis: error magnification near NCD–MD transitions} \label{sec:error_analyasis_generative}

\begin{figure}[t!]
    \centering
    \includegraphics[width=.54\textwidth]{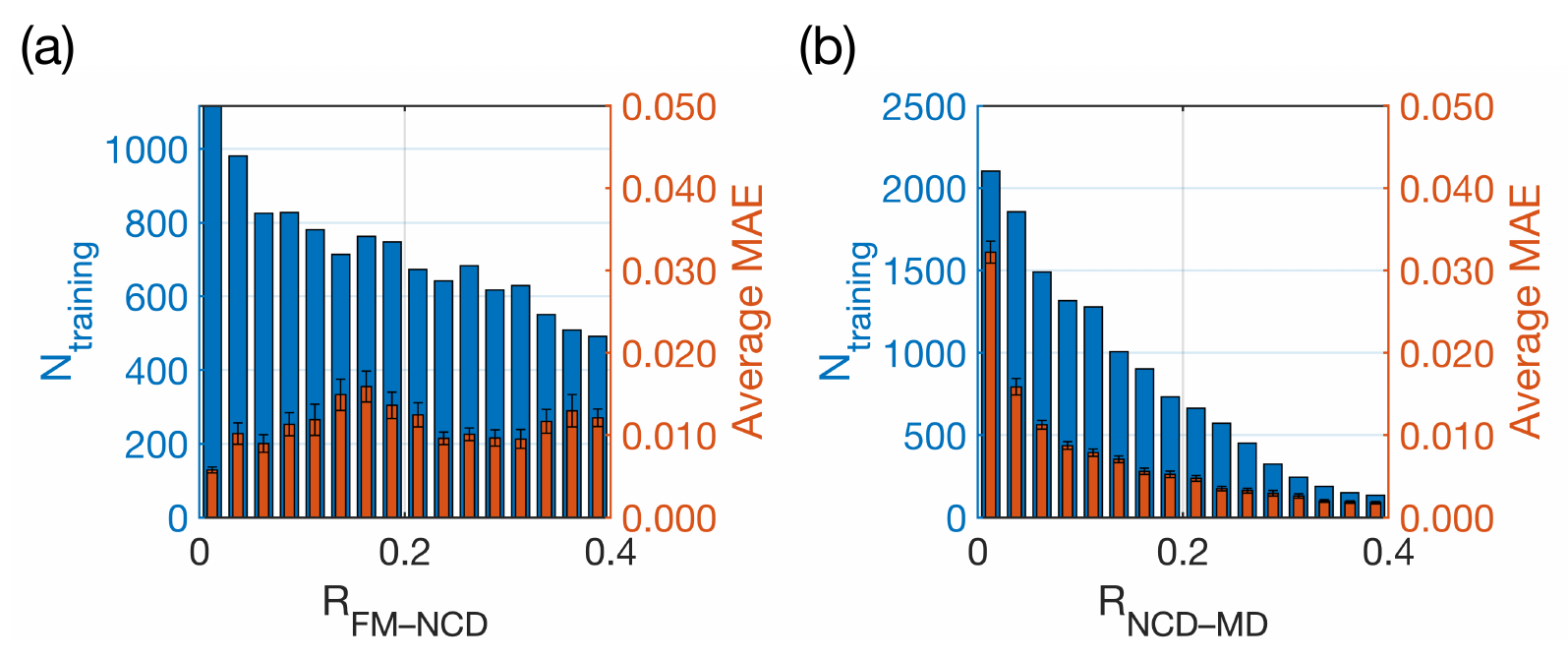}
    \caption[Error in generative]{Profiles of domain-image-prediction errors near phase boundaries. Blue bars depict the sample count in the training set. The $x$-axis indicates the Euclidean distance from each data point to the phase boundary. Red bars represent the average MAE values for out-of-plane magnetization between true and predicted images. Error bars signify the standard errors of the average MAE values.} 
    \label{fig9}
\end{figure}

To delve deeper into the errors in domain image generation, we evaluate the network's performance near phase boundaries. Figure~\ref{fig9} illustrates the profiles of average MAE values between true and predicted images concerning the Euclidean distance from each data point to the phase boundaries. We observe no discernible correlation between the average MAE value and the distance to the FM–NCD phase boundary (Fig.~\ref{fig9}(a)). However, the average MAE value demonstrates a striking correlation with the distance to the NCD–MD phase boundary, showing a steep increase near the phase boundary (Fig.~\ref{fig9}(b)). This trend is particularly compelling, given the rising number of samples within this region, and indicates that the nature of errors is deeply rooted in the physical aspects of the transition rather than a mere statistical origin, such as sample scarcity. These observations strongly suggest that the NCD–MD transition serves as a primary factor magnifying the errors in domain image generation, warranting further investigation.

One conceivable scenario is that the discontinuous nature of the NCD–MD transition exacerbates errors near this boundary. During this transition, a magnetic domain array undergoes a discontinuous transformation \cite{Hejazi10721, Zheng2022, Kim2023}. The degree of change, quantified by its net magnetization, tends to increase with a higher value of $A$ \cite{Kim2023}. In a large $A$ regime, this change may surpass the network's capacity to effectively capture, leading to the magnification of errors. This scenario gains support when observing the magnification of errors with increasing $A$ (Fig.~\ref{fig8}(e)). Consequently, it becomes evident that meticulous attention is required to mitigate the impacts of the abrupt change across the NCD–MD transition and uphold the accuracy of the generative model in this scenario.

\subsubsection{Prediction performance in averaged magnetization}

\begin{figure}[t!]
    \centering
    \includegraphics[width=.8\textwidth]{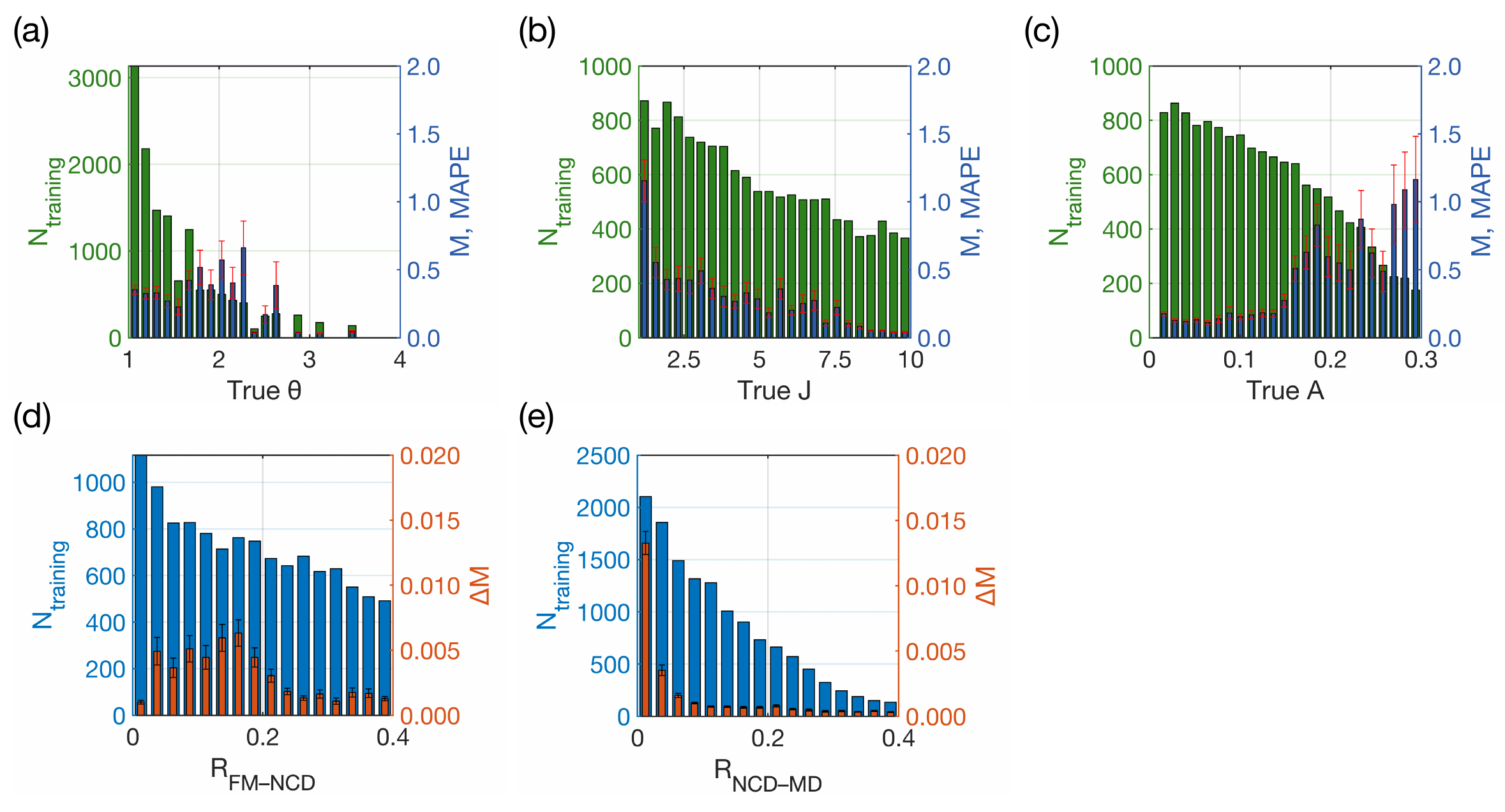}
    \caption{Performance of the trained network in the generative model in predicting averaged magnetization. (a–c) Profiles of MAPE values of averaged magnetization ($M$). The $x$-axis denotes the true values used in the simulation. Red bars represent the MAPE values for averaged magnetization. Error bars signify the standard errors of the MAPE values. (d–e) Profiles of MAE values of averaged magnetization near phase boundaries. The $x$-axis indicates the Euclidean distance from each data point to the phase boundary. Red bars represent the MAE values for averaged magnetization ($\Delta M$). Error bars signify the standard errors of the MAE values.}
    \label{fig10}
\end{figure}

To assess the versatility of the generative model, we evaluate its performance in predicting averaged magnetization, a practical indicator for discerning nanoscale magnetic domain arrays in experimental investigations \cite{Song2021, Xie2022, Xu2022, Xie2023, Cheng2023}. Figure~\ref{fig10}(a–c) presents the profiles of MAPE values between true and predicted averaged magnetization ($M$), concerning the parameter variations. Our findings reveal consistently low MAPE values across all parameter ranges. Particularly noteworthy is the observation that MAPE values remain below $0.3$ in the small twist angle regime ($\theta \lesssim 1.5 $\textdegree{}) (Fig.~\ref{fig10}(a)) and below $0.5$ near $(J, A) \approx (2.0\textrm{meV}, 0.2~\textrm{meV})$ pertinent to CrI\textsubscript{3} (Fig.~\ref{fig10}(b–c)). This capability indicates the network's proficiency in estimating averaged magnetization, underscoring its promising applications to small twist angle systems and the pivotal material CrI\textsubscript{3}, which are frequently investigated experimentally \cite{Song2021, Xie2022, Xu2022, Xie2023, Cheng2023}.

To delve deeper into the errors, we evaluate the network's performance near the phase boundaries. Figure~\ref{fig10}(d–e) illustrates the profiles of MAE values between true and predicted averaged magnetization ($\Delta M$), concerning the Euclidean distance to the phase boundaries. We observe no discernible correlation between $\Delta M$ and the distance to the FM–NCD phase boundary (Fig.~\ref{fig10}(d)). On the other hand, a striking increasing tendency of $\Delta M$ emerges near the NCD–MD phase boundary (Fig.~\ref{fig10}(e)). These observations align completely with the previous findings in domain image prediction of the generative model. Consequently, the considerations regarding error magnifying factors in that context apply similarly to the case of averaged magnetization, suggesting that the discontinuous nature of the NCD–MD transition serves as a primary factor in magnifying errors in predicting averaged magnetization.

\section{Discussion}

We have demonstrated the efficacy of our trained networks in accurately predicting magnetic Hamiltonian parameters based on magnetic domain images within twisted bilayer magnets and vice versa. The reliable performance of our DNN methods positions them as efficient tools for analyzing the intricate magnetic behaviors inherent in these systems. Specifically, these methods have direct applications in the study of twisted bilayer CrI\textsubscript{3} \cite{Song2021, Xu2022, Hejazi10721, PhysRevResearch.3.013027, Akram2021, Ghader2022, Zheng2022, Fumega_2023, Kim2023, Yang2023, PhysRevB.104.014410} and other related systems incorporating different van der Waals magnetic materials \cite{Hejazi10721, Akram2021, Fumega_2023, Kim2023, Akram2024}. Furthermore, our methods possess proficiency in the efficient analysis of twisted multilayer superlattice structures, which typically require increased computing resources for handling such large-sized superlattices \cite{Xie2022, Xie2023, Cheng2023, PhysRevB.108.L100401}. Additionally, the networks' robust performance against noise injections suggests promising applications to experimental data \cite{Song2021, Xu2022, Xie2022, Xie2023, Cheng2023}.

Future research prospects encompass expanding our methods to different moiré-induced magnetic orders in various twisted magnetic systems, including the exploration of the coexisting multiple zigzag antiferromagnetic orders in twisted bilayer \textalpha-RuCl\textsubscript{3} \cite{Akram2024}. Additionally, incorporation of topological defects such as skyrmions \cite{Tong2018, PhysRevB.103.L140406, PhysRevResearch.3.013027, PhysRevB.104.014410, Akram2021, Zheng2022, Ghader2022, PhysRevB.108.174440, Kim2023, Akram2024} and merons \cite{Kim2024} into the magnetic domain images holds promise for gaining valuable insights into the generation of topological defects within twisted van der Waals magnets, a captivating aspect that remains unexplored in this study.

We highlight that the pioneering deep-learning approach established in this study opens avenues for the application of deep-learning techniques in twisted van der Waals magnets. This significant breakthrough not only deepens our understanding of the intricate magnetic behaviors within these captivating systems but also holds great promise for accelerating advancements in this swiftly evolving field. \\

\section*{Acknowledgements}
We would like to thank Hyun Kim and Se Kwon Kim for sharing their insights. K.-M.K. was supported by the Institute for Basic Science in the Republic of Korea through the project IBS-R024-D1. T.S. was supported by the National Research Foundation of Korea (NRF) grant funded by the Korea government (MSIT) (2022R1F1A1074054).

\bibliography{ref.bib}

\begin{thebibliography}{42}%
\makeatletter
\providecommand \@ifxundefined [1]{%
 \@ifx{#1\undefined}
}%
\providecommand \@ifnum [1]{%
 \ifnum #1\expandafter \@firstoftwo
 \else \expandafter \@secondoftwo
 \fi
}%
\providecommand \@ifx [1]{%
 \ifx #1\expandafter \@firstoftwo
 \else \expandafter \@secondoftwo
 \fi
}%
\providecommand \natexlab [1]{#1}%
\providecommand \enquote  [1]{``#1''}%
\providecommand \bibnamefont  [1]{#1}%
\providecommand \bibfnamefont [1]{#1}%
\providecommand \citenamefont [1]{#1}%
\providecommand \href@noop [0]{\@secondoftwo}%
\providecommand \href [0]{\begingroup \@sanitize@url \@href}%
\providecommand \@href[1]{\@@startlink{#1}\@@href}%
\providecommand \@@href[1]{\endgroup#1\@@endlink}%
\providecommand \@sanitize@url [0]{\catcode `\\12\catcode `\$12\catcode
  `\&12\catcode `\#12\catcode `\^12\catcode `\_12\catcode `\%12\relax}%
\providecommand \@@startlink[1]{}%
\providecommand \@@endlink[0]{}%
\providecommand \url  [0]{\begingroup\@sanitize@url \@url }%
\providecommand \@url [1]{\endgroup\@href {#1}{\urlprefix }}%
\providecommand \urlprefix  [0]{URL }%
\providecommand \Eprint [0]{\href }%
\providecommand \doibase [0]{https://doi.org/}%
\providecommand \selectlanguage [0]{\@gobble}%
\providecommand \bibinfo  [0]{\@secondoftwo}%
\providecommand \bibfield  [0]{\@secondoftwo}%
\providecommand \translation [1]{[#1]}%
\providecommand \BibitemOpen [0]{}%
\providecommand \bibitemStop [0]{}%
\providecommand \bibitemNoStop [0]{.\EOS\space}%
\providecommand \EOS [0]{\spacefactor3000\relax}%
\providecommand \BibitemShut  [1]{\csname bibitem#1\endcsname}%
\let\auto@bib@innerbib\@empty
\bibitem [{\citenamefont {Song}\ \emph {et~al.}(2021)\citenamefont {Song},
  \citenamefont {Sun}, \citenamefont {Anderson}, \citenamefont {Wang},
  \citenamefont {Qian}, \citenamefont {Taniguchi}, \citenamefont {Watanabe},
  \citenamefont {McGuire}, \citenamefont {Stöhr}, \citenamefont {Xiao},
  \citenamefont {Cao}, \citenamefont {Wrachtrup},\ and\ \citenamefont
  {Xu}}]{Song2021}%
  \BibitemOpen
  \bibfield  {author} {\bibinfo {author} {\bibfnamefont {T.}~\bibnamefont
  {Song}}, \bibinfo {author} {\bibfnamefont {Q.-C.}\ \bibnamefont {Sun}},
  \bibinfo {author} {\bibfnamefont {E.}~\bibnamefont {Anderson}}, \bibinfo
  {author} {\bibfnamefont {C.}~\bibnamefont {Wang}}, \bibinfo {author}
  {\bibfnamefont {J.}~\bibnamefont {Qian}}, \bibinfo {author} {\bibfnamefont
  {T.}~\bibnamefont {Taniguchi}}, \bibinfo {author} {\bibfnamefont
  {K.}~\bibnamefont {Watanabe}}, \bibinfo {author} {\bibfnamefont {M.~A.}\
  \bibnamefont {McGuire}}, \bibinfo {author} {\bibfnamefont {R.}~\bibnamefont
  {Stöhr}}, \bibinfo {author} {\bibfnamefont {D.}~\bibnamefont {Xiao}},
  \bibinfo {author} {\bibfnamefont {T.}~\bibnamefont {Cao}}, \bibinfo {author}
  {\bibfnamefont {J.}~\bibnamefont {Wrachtrup}},\ and\ \bibinfo {author}
  {\bibfnamefont {X.}~\bibnamefont {Xu}},\ }\bibfield  {title} {\bibinfo
  {title} {Direct visualization of magnetic domains and moiré magnetism in
  twisted 2{D} magnets},\ }\href {https://doi.org/10.1126/science.abj7478}
  {\bibfield  {journal} {\bibinfo  {journal} {Science}\ }\textbf {\bibinfo
  {volume} {374}},\ \bibinfo {pages} {1140} (\bibinfo {year}
  {2021})}\BibitemShut {NoStop}%
\bibitem [{\citenamefont {Xu}\ \emph {et~al.}(2022)\citenamefont {Xu},
  \citenamefont {Ray}, \citenamefont {Shao}, \citenamefont {Jiang},
  \citenamefont {Lee}, \citenamefont {Weber}, \citenamefont {Goldberger},
  \citenamefont {Watanabe}, \citenamefont {Taniguchi}, \citenamefont {Muller},
  \citenamefont {Mak},\ and\ \citenamefont {Shan}}]{Xu2022}%
  \BibitemOpen
  \bibfield  {author} {\bibinfo {author} {\bibfnamefont {Y.}~\bibnamefont
  {Xu}}, \bibinfo {author} {\bibfnamefont {A.}~\bibnamefont {Ray}}, \bibinfo
  {author} {\bibfnamefont {Y.-T.}\ \bibnamefont {Shao}}, \bibinfo {author}
  {\bibfnamefont {S.}~\bibnamefont {Jiang}}, \bibinfo {author} {\bibfnamefont
  {K.}~\bibnamefont {Lee}}, \bibinfo {author} {\bibfnamefont {D.}~\bibnamefont
  {Weber}}, \bibinfo {author} {\bibfnamefont {J.~E.}\ \bibnamefont
  {Goldberger}}, \bibinfo {author} {\bibfnamefont {K.}~\bibnamefont
  {Watanabe}}, \bibinfo {author} {\bibfnamefont {T.}~\bibnamefont {Taniguchi}},
  \bibinfo {author} {\bibfnamefont {D.~A.}\ \bibnamefont {Muller}}, \bibinfo
  {author} {\bibfnamefont {K.~F.}\ \bibnamefont {Mak}},\ and\ \bibinfo {author}
  {\bibfnamefont {J.}~\bibnamefont {Shan}},\ }\bibfield  {title} {\bibinfo
  {title} {Coexisting ferromagnetic--antiferromagnetic state in twisted bilayer
  {C}r{I}\textsubscript{3}},\ }\href
  {https://doi.org/10.1038/s41565-021-01014-y} {\bibfield  {journal} {\bibinfo
  {journal} {Nat. Nanotechnol.}\ }\textbf {\bibinfo {volume} {17}},\ \bibinfo
  {pages} {143} (\bibinfo {year} {2022})}\BibitemShut {NoStop}%
\bibitem [{\citenamefont {Xie}\ \emph {et~al.}(2022)\citenamefont {Xie},
  \citenamefont {Luo}, \citenamefont {Ye}, \citenamefont {Ye}, \citenamefont
  {Ge}, \citenamefont {Sung}, \citenamefont {Rennich}, \citenamefont {Yan},
  \citenamefont {Fu}, \citenamefont {Tian}, \citenamefont {Lei}, \citenamefont
  {Hovden}, \citenamefont {Sun}, \citenamefont {He},\ and\ \citenamefont
  {Zhao}}]{Xie2022}%
  \BibitemOpen
  \bibfield  {author} {\bibinfo {author} {\bibfnamefont {H.}~\bibnamefont
  {Xie}}, \bibinfo {author} {\bibfnamefont {X.}~\bibnamefont {Luo}}, \bibinfo
  {author} {\bibfnamefont {G.}~\bibnamefont {Ye}}, \bibinfo {author}
  {\bibfnamefont {Z.}~\bibnamefont {Ye}}, \bibinfo {author} {\bibfnamefont
  {H.}~\bibnamefont {Ge}}, \bibinfo {author} {\bibfnamefont {S.~H.}\
  \bibnamefont {Sung}}, \bibinfo {author} {\bibfnamefont {E.}~\bibnamefont
  {Rennich}}, \bibinfo {author} {\bibfnamefont {S.}~\bibnamefont {Yan}},
  \bibinfo {author} {\bibfnamefont {Y.}~\bibnamefont {Fu}}, \bibinfo {author}
  {\bibfnamefont {S.}~\bibnamefont {Tian}}, \bibinfo {author} {\bibfnamefont
  {H.}~\bibnamefont {Lei}}, \bibinfo {author} {\bibfnamefont {R.}~\bibnamefont
  {Hovden}}, \bibinfo {author} {\bibfnamefont {K.}~\bibnamefont {Sun}},
  \bibinfo {author} {\bibfnamefont {R.}~\bibnamefont {He}},\ and\ \bibinfo
  {author} {\bibfnamefont {L.}~\bibnamefont {Zhao}},\ }\bibfield  {title}
  {\bibinfo {title} {Twist engineering of the two-dimensional magnetism in
  double bilayer chromium triiodide homostructures},\ }\href
  {https://doi.org/10.1038/s41567-021-01408-8} {\bibfield  {journal} {\bibinfo
  {journal} {Nat. Phys.}\ }\textbf {\bibinfo {volume} {18}},\ \bibinfo {pages}
  {30} (\bibinfo {year} {2022})}\BibitemShut {NoStop}%
\bibitem [{\citenamefont {Xie}\ \emph {et~al.}(2023)\citenamefont {Xie},
  \citenamefont {Luo}, \citenamefont {Ye}, \citenamefont {Sun}, \citenamefont
  {Ye}, \citenamefont {Sung}, \citenamefont {Ge}, \citenamefont {Yan},
  \citenamefont {Fu}, \citenamefont {Tian}, \citenamefont {Lei}, \citenamefont
  {Sun}, \citenamefont {Hovden}, \citenamefont {He},\ and\ \citenamefont
  {Zhao}}]{Xie2023}%
  \BibitemOpen
  \bibfield  {author} {\bibinfo {author} {\bibfnamefont {H.}~\bibnamefont
  {Xie}}, \bibinfo {author} {\bibfnamefont {X.}~\bibnamefont {Luo}}, \bibinfo
  {author} {\bibfnamefont {Z.}~\bibnamefont {Ye}}, \bibinfo {author}
  {\bibfnamefont {Z.}~\bibnamefont {Sun}}, \bibinfo {author} {\bibfnamefont
  {G.}~\bibnamefont {Ye}}, \bibinfo {author} {\bibfnamefont {S.~H.}\
  \bibnamefont {Sung}}, \bibinfo {author} {\bibfnamefont {H.}~\bibnamefont
  {Ge}}, \bibinfo {author} {\bibfnamefont {S.}~\bibnamefont {Yan}}, \bibinfo
  {author} {\bibfnamefont {Y.}~\bibnamefont {Fu}}, \bibinfo {author}
  {\bibfnamefont {S.}~\bibnamefont {Tian}}, \bibinfo {author} {\bibfnamefont
  {H.}~\bibnamefont {Lei}}, \bibinfo {author} {\bibfnamefont {K.}~\bibnamefont
  {Sun}}, \bibinfo {author} {\bibfnamefont {R.}~\bibnamefont {Hovden}},
  \bibinfo {author} {\bibfnamefont {R.}~\bibnamefont {He}},\ and\ \bibinfo
  {author} {\bibfnamefont {L.}~\bibnamefont {Zhao}},\ }\bibfield  {title}
  {\bibinfo {title} {Evidence of non-collinear spin texture in magnetic
  moir{\'e} superlattices},\ }\href
  {https://doi.org/10.1038/s41567-023-02061-z} {\bibfield  {journal} {\bibinfo
  {journal} {Nat. Phys.}\ }\textbf {\bibinfo {volume} {19}},\ \bibinfo {pages}
  {1150} (\bibinfo {year} {2023})}\BibitemShut {NoStop}%
\bibitem [{\citenamefont {Cheng}\ \emph {et~al.}(2023)\citenamefont {Cheng},
  \citenamefont {Rahman}, \citenamefont {Allcca}, \citenamefont {Rustagi},
  \citenamefont {Liu}, \citenamefont {Liu}, \citenamefont {Fu}, \citenamefont
  {Zhu}, \citenamefont {Mao}, \citenamefont {Watanabe}, \citenamefont
  {Taniguchi}, \citenamefont {Upadhyaya},\ and\ \citenamefont
  {Chen}}]{Cheng2023}%
  \BibitemOpen
  \bibfield  {author} {\bibinfo {author} {\bibfnamefont {G.}~\bibnamefont
  {Cheng}}, \bibinfo {author} {\bibfnamefont {M.~M.}\ \bibnamefont {Rahman}},
  \bibinfo {author} {\bibfnamefont {A.~L.}\ \bibnamefont {Allcca}}, \bibinfo
  {author} {\bibfnamefont {A.}~\bibnamefont {Rustagi}}, \bibinfo {author}
  {\bibfnamefont {X.}~\bibnamefont {Liu}}, \bibinfo {author} {\bibfnamefont
  {L.}~\bibnamefont {Liu}}, \bibinfo {author} {\bibfnamefont {L.}~\bibnamefont
  {Fu}}, \bibinfo {author} {\bibfnamefont {Y.}~\bibnamefont {Zhu}}, \bibinfo
  {author} {\bibfnamefont {Z.}~\bibnamefont {Mao}}, \bibinfo {author}
  {\bibfnamefont {K.}~\bibnamefont {Watanabe}}, \bibinfo {author}
  {\bibfnamefont {T.}~\bibnamefont {Taniguchi}}, \bibinfo {author}
  {\bibfnamefont {P.}~\bibnamefont {Upadhyaya}},\ and\ \bibinfo {author}
  {\bibfnamefont {Y.~P.}\ \bibnamefont {Chen}},\ }\bibfield  {title} {\bibinfo
  {title} {Electrically tunable moir{\'e} magnetism in twisted double bilayers
  of chromium triiodide},\ }\href {https://doi.org/10.1038/s41928-023-00978-0}
  {\bibfield  {journal} {\bibinfo  {journal} {Nat. Electron.}\ }\textbf
  {\bibinfo {volume} {6}},\ \bibinfo {pages} {434} (\bibinfo {year}
  {2023})}\BibitemShut {NoStop}%
\bibitem [{\citenamefont {Hejazi}\ \emph {et~al.}(2020)\citenamefont {Hejazi},
  \citenamefont {Luo},\ and\ \citenamefont {Balents}}]{Hejazi10721}%
  \BibitemOpen
  \bibfield  {author} {\bibinfo {author} {\bibfnamefont {K.}~\bibnamefont
  {Hejazi}}, \bibinfo {author} {\bibfnamefont {Z.-X.}\ \bibnamefont {Luo}},\
  and\ \bibinfo {author} {\bibfnamefont {L.}~\bibnamefont {Balents}},\
  }\bibfield  {title} {\bibinfo {title} {Noncollinear phases in moir{\'e}
  magnets},\ }\href {https://doi.org/10.1073/pnas.2000347117} {\bibfield
  {journal} {\bibinfo  {journal} {Proc. Natl. Acad. Sci. U.S.A.}\ }\textbf
  {\bibinfo {volume} {117}},\ \bibinfo {pages} {10721} (\bibinfo {year}
  {2020})}\BibitemShut {NoStop}%
\bibitem [{\citenamefont {Akram}\ \emph {et~al.}(2021)\citenamefont {Akram},
  \citenamefont {LaBollita}, \citenamefont {Dey}, \citenamefont {Kapeghian},
  \citenamefont {Erten},\ and\ \citenamefont {Botana}}]{Akram2021}%
  \BibitemOpen
  \bibfield  {author} {\bibinfo {author} {\bibfnamefont {M.}~\bibnamefont
  {Akram}}, \bibinfo {author} {\bibfnamefont {H.}~\bibnamefont {LaBollita}},
  \bibinfo {author} {\bibfnamefont {D.}~\bibnamefont {Dey}}, \bibinfo {author}
  {\bibfnamefont {J.}~\bibnamefont {Kapeghian}}, \bibinfo {author}
  {\bibfnamefont {O.}~\bibnamefont {Erten}},\ and\ \bibinfo {author}
  {\bibfnamefont {A.~S.}\ \bibnamefont {Botana}},\ }\bibfield  {title}
  {\bibinfo {title} {Moir{\'e} skyrmions and chiral magnetic phases in twisted
  {C}r{X}\textsubscript{3} ({X} = {I}, {B}r, and {C}l) bilayers},\ }\href
  {https://doi.org/10.1021/acs.nanolett.1c02096} {\bibfield  {journal}
  {\bibinfo  {journal} {Nano Lett.}\ }\textbf {\bibinfo {volume} {21}},\
  \bibinfo {pages} {6633} (\bibinfo {year} {2021})}\BibitemShut {NoStop}%
\bibitem [{\citenamefont {Zheng}(2023)}]{Zheng2022}%
  \BibitemOpen
  \bibfield  {author} {\bibinfo {author} {\bibfnamefont {F.}~\bibnamefont
  {Zheng}},\ }\bibfield  {title} {\bibinfo {title} {Magnetic skyrmion lattices
  in a novel 2{D}-twisted bilayer magnet},\ }\href
  {https://doi.org/https://doi.org/10.1002/adfm.202206923} {\bibfield
  {journal} {\bibinfo  {journal} {Adv. Func. Mater.}\ }\textbf {\bibinfo
  {volume} {33}},\ \bibinfo {pages} {2206923} (\bibinfo {year}
  {2023})}\BibitemShut {NoStop}%
\bibitem [{\citenamefont {Kim}\ \emph {et~al.}(2023)\citenamefont {Kim},
  \citenamefont {Kiem}, \citenamefont {Bednik}, \citenamefont {Han},\ and\
  \citenamefont {Park}}]{Kim2023}%
  \BibitemOpen
  \bibfield  {author} {\bibinfo {author} {\bibfnamefont {K.-M.}\ \bibnamefont
  {Kim}}, \bibinfo {author} {\bibfnamefont {D.~H.}\ \bibnamefont {Kiem}},
  \bibinfo {author} {\bibfnamefont {G.}~\bibnamefont {Bednik}}, \bibinfo
  {author} {\bibfnamefont {M.~J.}\ \bibnamefont {Han}},\ and\ \bibinfo {author}
  {\bibfnamefont {M.~J.}\ \bibnamefont {Park}},\ }\bibfield  {title} {\bibinfo
  {title} {Ab initio spin hamiltonian and topological noncentrosymmetric
  magnetism in twisted bilayer cri3},\ }\href
  {https://doi.org/10.1021/acs.nanolett.3c01529} {\bibfield  {journal}
  {\bibinfo  {journal} {Nano Lett.}\ }\textbf {\bibinfo {volume} {23}},\
  \bibinfo {pages} {6088} (\bibinfo {year} {2023})}\BibitemShut {NoStop}%
\bibitem [{\citenamefont {Yang}\ \emph {et~al.}(2023)\citenamefont {Yang},
  \citenamefont {Li}, \citenamefont {Xiang}, \citenamefont {Lin},\ and\
  \citenamefont {Huang}}]{Yang2023}%
  \BibitemOpen
  \bibfield  {author} {\bibinfo {author} {\bibfnamefont {B.}~\bibnamefont
  {Yang}}, \bibinfo {author} {\bibfnamefont {Y.}~\bibnamefont {Li}}, \bibinfo
  {author} {\bibfnamefont {H.}~\bibnamefont {Xiang}}, \bibinfo {author}
  {\bibfnamefont {H.}~\bibnamefont {Lin}},\ and\ \bibinfo {author}
  {\bibfnamefont {B.}~\bibnamefont {Huang}},\ }\bibfield  {title} {\bibinfo
  {title} {Moir{\'e} magnetic exchange interactions in twisted magnets},\
  }\href {https://doi.org/10.1038/s43588-023-00430-5} {\bibfield  {journal}
  {\bibinfo  {journal} {Nat. Comput. Sci.}\ }\textbf {\bibinfo {volume} {3}},\
  \bibinfo {pages} {314} (\bibinfo {year} {2023})}\BibitemShut {NoStop}%
\bibitem [{\citenamefont {Hejazi}\ \emph {et~al.}(2021)\citenamefont {Hejazi},
  \citenamefont {Luo},\ and\ \citenamefont {Balents}}]{PhysRevB.104.L100406}%
  \BibitemOpen
  \bibfield  {author} {\bibinfo {author} {\bibfnamefont {K.}~\bibnamefont
  {Hejazi}}, \bibinfo {author} {\bibfnamefont {Z.-X.}\ \bibnamefont {Luo}},\
  and\ \bibinfo {author} {\bibfnamefont {L.}~\bibnamefont {Balents}},\
  }\bibfield  {title} {\bibinfo {title} {Heterobilayer moir\'e magnets: Moir\'e
  skyrmions and commensurate-incommensurate transitions},\ }\href
  {https://doi.org/10.1103/PhysRevB.104.L100406} {\bibfield  {journal}
  {\bibinfo  {journal} {Phys. Rev. B}\ }\textbf {\bibinfo {volume} {104}},\
  \bibinfo {pages} {L100406} (\bibinfo {year} {2021})}\BibitemShut {NoStop}%
\bibitem [{\citenamefont {Shaban}\ \emph {et~al.}(2023)\citenamefont {Shaban},
  \citenamefont {Lobanov}, \citenamefont {Uzdin},\ and\ \citenamefont
  {Iorsh}}]{PhysRevB.108.174440}%
  \BibitemOpen
  \bibfield  {author} {\bibinfo {author} {\bibfnamefont {P.~S.}\ \bibnamefont
  {Shaban}}, \bibinfo {author} {\bibfnamefont {I.~S.}\ \bibnamefont {Lobanov}},
  \bibinfo {author} {\bibfnamefont {V.~M.}\ \bibnamefont {Uzdin}},\ and\
  \bibinfo {author} {\bibfnamefont {I.~V.}\ \bibnamefont {Iorsh}},\ }\bibfield
  {title} {\bibinfo {title} {Skyrmion dynamics in moir\'e magnets},\ }\href
  {https://doi.org/10.1103/PhysRevB.108.174440} {\bibfield  {journal} {\bibinfo
   {journal} {Phys. Rev. B}\ }\textbf {\bibinfo {volume} {108}},\ \bibinfo
  {pages} {174440} (\bibinfo {year} {2023})}\BibitemShut {NoStop}%
\bibitem [{\citenamefont {Tong}\ \emph {et~al.}(2018)\citenamefont {Tong},
  \citenamefont {Liu}, \citenamefont {Xiao},\ and\ \citenamefont
  {Yao}}]{Tong2018}%
  \BibitemOpen
  \bibfield  {author} {\bibinfo {author} {\bibfnamefont {Q.}~\bibnamefont
  {Tong}}, \bibinfo {author} {\bibfnamefont {F.}~\bibnamefont {Liu}}, \bibinfo
  {author} {\bibfnamefont {J.}~\bibnamefont {Xiao}},\ and\ \bibinfo {author}
  {\bibfnamefont {W.}~\bibnamefont {Yao}},\ }\bibfield  {title} {\bibinfo
  {title} {Skyrmions in the moiré of van der {W}aals 2{D} magnets},\ }\href
  {https://doi.org/10.1021/acs.nanolett.8b03315} {\bibfield  {journal}
  {\bibinfo  {journal} {Nano Lett.}\ }\textbf {\bibinfo {volume} {18}},\
  \bibinfo {pages} {7194} (\bibinfo {year} {2018})}\BibitemShut {NoStop}%
\bibitem [{\citenamefont {Akram}\ and\ \citenamefont
  {Erten}(2021)}]{PhysRevB.103.L140406}%
  \BibitemOpen
  \bibfield  {author} {\bibinfo {author} {\bibfnamefont {M.}~\bibnamefont
  {Akram}}\ and\ \bibinfo {author} {\bibfnamefont {O.}~\bibnamefont {Erten}},\
  }\bibfield  {title} {\bibinfo {title} {Skyrmions in twisted van der waals
  magnets},\ }\href {https://doi.org/10.1103/PhysRevB.103.L140406} {\bibfield
  {journal} {\bibinfo  {journal} {Phys. Rev. B}\ }\textbf {\bibinfo {volume}
  {103}},\ \bibinfo {pages} {L140406} (\bibinfo {year} {2021})}\BibitemShut
  {NoStop}%
\bibitem [{\citenamefont {Ray}\ and\ \citenamefont
  {Das}(2021)}]{PhysRevB.104.014410}%
  \BibitemOpen
  \bibfield  {author} {\bibinfo {author} {\bibfnamefont {S.}~\bibnamefont
  {Ray}}\ and\ \bibinfo {author} {\bibfnamefont {T.}~\bibnamefont {Das}},\
  }\bibfield  {title} {\bibinfo {title} {Hierarchy of multi-order skyrmion
  phases in twisted magnetic bilayers},\ }\href
  {https://doi.org/10.1103/PhysRevB.104.014410} {\bibfield  {journal} {\bibinfo
   {journal} {Phys. Rev. B}\ }\textbf {\bibinfo {volume} {104}},\ \bibinfo
  {pages} {014410} (\bibinfo {year} {2021})}\BibitemShut {NoStop}%
\bibitem [{\citenamefont {Xiao}\ \emph {et~al.}(2021)\citenamefont {Xiao},
  \citenamefont {Chen},\ and\ \citenamefont {Tong}}]{PhysRevResearch.3.013027}%
  \BibitemOpen
  \bibfield  {author} {\bibinfo {author} {\bibfnamefont {F.}~\bibnamefont
  {Xiao}}, \bibinfo {author} {\bibfnamefont {K.}~\bibnamefont {Chen}},\ and\
  \bibinfo {author} {\bibfnamefont {Q.}~\bibnamefont {Tong}},\ }\bibfield
  {title} {\bibinfo {title} {Magnetization textures in twisted bilayer
  {C}r{X}\textsubscript{3} ({X}={B}r, {I})},\ }\href
  {https://doi.org/10.1103/PhysRevResearch.3.013027} {\bibfield  {journal}
  {\bibinfo  {journal} {Phys. Rev. Res.}\ }\textbf {\bibinfo {volume} {3}},\
  \bibinfo {pages} {013027} (\bibinfo {year} {2021})}\BibitemShut {NoStop}%
\bibitem [{\citenamefont {Ghader}\ \emph {et~al.}(2022)\citenamefont {Ghader},
  \citenamefont {Jabakhanji},\ and\ \citenamefont {Stroppa}}]{Ghader2022}%
  \BibitemOpen
  \bibfield  {author} {\bibinfo {author} {\bibfnamefont {D.}~\bibnamefont
  {Ghader}}, \bibinfo {author} {\bibfnamefont {B.}~\bibnamefont {Jabakhanji}},\
  and\ \bibinfo {author} {\bibfnamefont {A.}~\bibnamefont {Stroppa}},\
  }\bibfield  {title} {\bibinfo {title} {Whirling interlayer fields as a source
  of stable topological order in moir{\'e} {C}r{I}\textsubscript{3}},\ }\href
  {https://doi.org/10.1038/s42005-022-00972-6} {\bibfield  {journal} {\bibinfo
  {journal} {Commun. Phys.}\ }\textbf {\bibinfo {volume} {5}},\ \bibinfo
  {pages} {192} (\bibinfo {year} {2022})}\BibitemShut {NoStop}%
\bibitem [{\citenamefont {Fumega}\ and\ \citenamefont
  {Lado}(2023)}]{Fumega_2023}%
  \BibitemOpen
  \bibfield  {author} {\bibinfo {author} {\bibfnamefont {A.~O.}\ \bibnamefont
  {Fumega}}\ and\ \bibinfo {author} {\bibfnamefont {J.~L.}\ \bibnamefont
  {Lado}},\ }\bibfield  {title} {\bibinfo {title} {Moiré-driven multiferroic
  order in twisted {C}r{C}l\textsubscript{3}, {C}r{B}r\textsubscript{3} and
  {C}r{I}\textsubscript{3} bilayers},\ }\href
  {https://doi.org/10.1088/2053-1583/acc671} {\bibfield  {journal} {\bibinfo
  {journal} {2D Materials}\ }\textbf {\bibinfo {volume} {10}},\ \bibinfo
  {pages} {025026} (\bibinfo {year} {2023})}\BibitemShut {NoStop}%
\bibitem [{\citenamefont {Kim}\ and\ \citenamefont
  {Park}(2023)}]{PhysRevB.108.L100401}%
  \BibitemOpen
  \bibfield  {author} {\bibinfo {author} {\bibfnamefont {K.-M.}\ \bibnamefont
  {Kim}}\ and\ \bibinfo {author} {\bibfnamefont {M.~J.}\ \bibnamefont {Park}},\
  }\bibfield  {title} {\bibinfo {title} {Controllable magnetic domains in
  twisted trilayer magnets},\ }\href
  {https://doi.org/10.1103/PhysRevB.108.L100401} {\bibfield  {journal}
  {\bibinfo  {journal} {Phys. Rev. B}\ }\textbf {\bibinfo {volume} {108}},\
  \bibinfo {pages} {L100401} (\bibinfo {year} {2023})}\BibitemShut {NoStop}%
\bibitem [{\citenamefont {Kim}\ \emph {et~al.}(2024)\citenamefont {Kim},
  \citenamefont {Go}, \citenamefont {Park},\ and\ \citenamefont
  {Kim}}]{Kim2024}%
  \BibitemOpen
  \bibfield  {author} {\bibinfo {author} {\bibfnamefont {K.-M.}\ \bibnamefont
  {Kim}}, \bibinfo {author} {\bibfnamefont {G.}~\bibnamefont {Go}}, \bibinfo
  {author} {\bibfnamefont {M.~J.}\ \bibnamefont {Park}},\ and\ \bibinfo
  {author} {\bibfnamefont {S.~K.}\ \bibnamefont {Kim}},\ }\bibfield  {title}
  {\bibinfo {title} {Emergence of stable meron quartets in twisted magnets},\
  }\href {https://doi.org/10.1021/acs.nanolett.3c03246} {\bibfield  {journal}
  {\bibinfo  {journal} {Nano Lett.}\ }\textbf {\bibinfo {volume} {24}},\
  \bibinfo {pages} {74} (\bibinfo {year} {2024})}\BibitemShut {NoStop}%
\bibitem [{\citenamefont {Akram}\ \emph {et~al.}(2024)\citenamefont {Akram},
  \citenamefont {Kapeghian}, \citenamefont {Das}, \citenamefont {Valent{\'i}},
  \citenamefont {Botana},\ and\ \citenamefont {Erten}}]{Akram2024}%
  \BibitemOpen
  \bibfield  {author} {\bibinfo {author} {\bibfnamefont {M.}~\bibnamefont
  {Akram}}, \bibinfo {author} {\bibfnamefont {J.}~\bibnamefont {Kapeghian}},
  \bibinfo {author} {\bibfnamefont {J.}~\bibnamefont {Das}}, \bibinfo {author}
  {\bibfnamefont {R.}~\bibnamefont {Valent{\'i}}}, \bibinfo {author}
  {\bibfnamefont {A.~S.}\ \bibnamefont {Botana}},\ and\ \bibinfo {author}
  {\bibfnamefont {O.}~\bibnamefont {Erten}},\ }\bibfield  {title} {\bibinfo
  {title} {Theory of moir{\'e} magnetism in twisted bilayer $\alpha$-rucl3},\
  }\href {https://doi.org/10.1021/acs.nanolett.3c04084} {\bibfield  {journal}
  {\bibinfo  {journal} {Nano Lett.}\ }\textbf {\bibinfo {volume} {24}},\
  \bibinfo {pages} {890} (\bibinfo {year} {2024})}\BibitemShut {NoStop}%
\bibitem [{\citenamefont {Kwon}\ \emph {et~al.}(2020)\citenamefont {Kwon},
  \citenamefont {Yoon}, \citenamefont {Lee}, \citenamefont {Chen},
  \citenamefont {Liu}, \citenamefont {Schmid}, \citenamefont {Wu},
  \citenamefont {Choi},\ and\ \citenamefont
  {Won}}]{doi:10.1126/sciadv.abb0872}%
  \BibitemOpen
  \bibfield  {author} {\bibinfo {author} {\bibfnamefont {H.~Y.}\ \bibnamefont
  {Kwon}}, \bibinfo {author} {\bibfnamefont {H.~G.}\ \bibnamefont {Yoon}},
  \bibinfo {author} {\bibfnamefont {C.}~\bibnamefont {Lee}}, \bibinfo {author}
  {\bibfnamefont {G.}~\bibnamefont {Chen}}, \bibinfo {author} {\bibfnamefont
  {K.}~\bibnamefont {Liu}}, \bibinfo {author} {\bibfnamefont {A.~K.}\
  \bibnamefont {Schmid}}, \bibinfo {author} {\bibfnamefont {Y.~Z.}\
  \bibnamefont {Wu}}, \bibinfo {author} {\bibfnamefont {J.~W.}\ \bibnamefont
  {Choi}},\ and\ \bibinfo {author} {\bibfnamefont {C.}~\bibnamefont {Won}},\
  }\bibfield  {title} {\bibinfo {title} {Magnetic hamiltonian parameter
  estimation using deep learning techniques},\ }\href
  {https://doi.org/10.1126/sciadv.abb0872} {\bibfield  {journal} {\bibinfo
  {journal} {Sci. Adv.}\ }\textbf {\bibinfo {volume} {6}},\ \bibinfo {pages}
  {eabb0872} (\bibinfo {year} {2020})}\BibitemShut {NoStop}%
\bibitem [{\citenamefont {Wright}\ \emph {et~al.}(2022)\citenamefont {Wright},
  \citenamefont {Onodera}, \citenamefont {Stein}, \citenamefont {Wang},
  \citenamefont {Schachter}, \citenamefont {Hu},\ and\ \citenamefont
  {McMahon}}]{Wright2022}%
  \BibitemOpen
  \bibfield  {author} {\bibinfo {author} {\bibfnamefont {L.~G.}\ \bibnamefont
  {Wright}}, \bibinfo {author} {\bibfnamefont {T.}~\bibnamefont {Onodera}},
  \bibinfo {author} {\bibfnamefont {M.~M.}\ \bibnamefont {Stein}}, \bibinfo
  {author} {\bibfnamefont {T.}~\bibnamefont {Wang}}, \bibinfo {author}
  {\bibfnamefont {D.~T.}\ \bibnamefont {Schachter}}, \bibinfo {author}
  {\bibfnamefont {Z.}~\bibnamefont {Hu}},\ and\ \bibinfo {author}
  {\bibfnamefont {P.~L.}\ \bibnamefont {McMahon}},\ }\bibfield  {title}
  {\bibinfo {title} {Deep physical neural networks trained with
  backpropagation},\ }\href {https://doi.org/10.1038/s41586-021-04223-6}
  {\bibfield  {journal} {\bibinfo  {journal} {Nature}\ }\textbf {\bibinfo
  {volume} {601}},\ \bibinfo {pages} {549} (\bibinfo {year}
  {2022})}\BibitemShut {NoStop}%
\bibitem [{\citenamefont {Schmidhuber}(2015)}]{SCHMIDHUBER201585}%
  \BibitemOpen
  \bibfield  {author} {\bibinfo {author} {\bibfnamefont {J.}~\bibnamefont
  {Schmidhuber}},\ }\bibfield  {title} {\bibinfo {title} {Deep learning in
  neural networks: An overview},\ }\href
  {https://doi.org/https://doi.org/10.1016/j.neunet.2014.09.003} {\bibfield
  {journal} {\bibinfo  {journal} {Neural Networks}\ }\textbf {\bibinfo {volume}
  {61}},\ \bibinfo {pages} {85} (\bibinfo {year} {2015})}\BibitemShut {NoStop}%
\bibitem [{\citenamefont {Pfau}\ \emph {et~al.}(2020)\citenamefont {Pfau},
  \citenamefont {Spencer}, \citenamefont {Matthews},\ and\ \citenamefont
  {Foulkes}}]{PhysRevResearch.2.033429}%
  \BibitemOpen
  \bibfield  {author} {\bibinfo {author} {\bibfnamefont {D.}~\bibnamefont
  {Pfau}}, \bibinfo {author} {\bibfnamefont {J.~S.}\ \bibnamefont {Spencer}},
  \bibinfo {author} {\bibfnamefont {A.~G. D.~G.}\ \bibnamefont {Matthews}},\
  and\ \bibinfo {author} {\bibfnamefont {W.~M.~C.}\ \bibnamefont {Foulkes}},\
  }\bibfield  {title} {\bibinfo {title} {Ab initio solution of the
  many-electron schr\"odinger equation with deep neural networks},\ }\href
  {https://doi.org/10.1103/PhysRevResearch.2.033429} {\bibfield  {journal}
  {\bibinfo  {journal} {Phys. Rev. Res.}\ }\textbf {\bibinfo {volume} {2}},\
  \bibinfo {pages} {033429} (\bibinfo {year} {2020})}\BibitemShut {NoStop}%
\bibitem [{\citenamefont {Senior}\ \emph {et~al.}(2019)\citenamefont {Senior},
  \citenamefont {Evans}, \citenamefont {Jumper}, \citenamefont {Kirkpatrick},
  \citenamefont {Sifre}, \citenamefont {Green}, \citenamefont {Qin},
  \citenamefont {Žídek}, \citenamefont {Nelson}, \citenamefont {Bridgland},
  \citenamefont {Penedones}, \citenamefont {Petersen}, \citenamefont
  {Simonyan}, \citenamefont {Crossan}, \citenamefont {Kohli}, \citenamefont
  {Jones}, \citenamefont {Silver}, \citenamefont {Kavukcuoglu},\ and\
  \citenamefont {Hassabis}}]{https://doi.org/10.1002/prot.25834}%
  \BibitemOpen
  \bibfield  {author} {\bibinfo {author} {\bibfnamefont {A.~W.}\ \bibnamefont
  {Senior}}, \bibinfo {author} {\bibfnamefont {R.}~\bibnamefont {Evans}},
  \bibinfo {author} {\bibfnamefont {J.}~\bibnamefont {Jumper}}, \bibinfo
  {author} {\bibfnamefont {J.}~\bibnamefont {Kirkpatrick}}, \bibinfo {author}
  {\bibfnamefont {L.}~\bibnamefont {Sifre}}, \bibinfo {author} {\bibfnamefont
  {T.}~\bibnamefont {Green}}, \bibinfo {author} {\bibfnamefont
  {C.}~\bibnamefont {Qin}}, \bibinfo {author} {\bibfnamefont {A.}~\bibnamefont
  {Žídek}}, \bibinfo {author} {\bibfnamefont {A.~W.~R.}\ \bibnamefont
  {Nelson}}, \bibinfo {author} {\bibfnamefont {A.}~\bibnamefont {Bridgland}},
  \bibinfo {author} {\bibfnamefont {H.}~\bibnamefont {Penedones}}, \bibinfo
  {author} {\bibfnamefont {S.}~\bibnamefont {Petersen}}, \bibinfo {author}
  {\bibfnamefont {K.}~\bibnamefont {Simonyan}}, \bibinfo {author}
  {\bibfnamefont {S.}~\bibnamefont {Crossan}}, \bibinfo {author} {\bibfnamefont
  {P.}~\bibnamefont {Kohli}}, \bibinfo {author} {\bibfnamefont {D.~T.}\
  \bibnamefont {Jones}}, \bibinfo {author} {\bibfnamefont {D.}~\bibnamefont
  {Silver}}, \bibinfo {author} {\bibfnamefont {K.}~\bibnamefont
  {Kavukcuoglu}},\ and\ \bibinfo {author} {\bibfnamefont {D.}~\bibnamefont
  {Hassabis}},\ }\bibfield  {title} {\bibinfo {title} {Protein structure
  prediction using multiple deep neural networks in the 13th critical
  assessment of protein structure prediction (casp13)},\ }\href
  {https://doi.org/https://doi.org/10.1002/prot.25834} {\bibfield  {journal}
  {\bibinfo  {journal} {Proteins: Structure, Function, and Bioinformatics}\
  }\textbf {\bibinfo {volume} {87}},\ \bibinfo {pages} {1141} (\bibinfo {year}
  {2019})}\BibitemShut {NoStop}%
\bibitem [{\citenamefont {Carleo}\ and\ \citenamefont
  {Troyer}(2017)}]{doi:10.1126/science.aag2302}%
  \BibitemOpen
  \bibfield  {author} {\bibinfo {author} {\bibfnamefont {G.}~\bibnamefont
  {Carleo}}\ and\ \bibinfo {author} {\bibfnamefont {M.}~\bibnamefont
  {Troyer}},\ }\bibfield  {title} {\bibinfo {title} {Solving the quantum
  many-body problem with artificial neural networks},\ }\href
  {https://doi.org/10.1126/science.aag2302} {\bibfield  {journal} {\bibinfo
  {journal} {Science}\ }\textbf {\bibinfo {volume} {355}},\ \bibinfo {pages}
  {602} (\bibinfo {year} {2017})}\BibitemShut {NoStop}%
\bibitem [{\citenamefont {Cai}\ and\ \citenamefont
  {Liu}(2018)}]{PhysRevB.97.035116}%
  \BibitemOpen
  \bibfield  {author} {\bibinfo {author} {\bibfnamefont {Z.}~\bibnamefont
  {Cai}}\ and\ \bibinfo {author} {\bibfnamefont {J.}~\bibnamefont {Liu}},\
  }\bibfield  {title} {\bibinfo {title} {Approximating quantum many-body wave
  functions using artificial neural networks},\ }\href
  {https://doi.org/10.1103/PhysRevB.97.035116} {\bibfield  {journal} {\bibinfo
  {journal} {Phys. Rev. B}\ }\textbf {\bibinfo {volume} {97}},\ \bibinfo
  {pages} {035116} (\bibinfo {year} {2018})}\BibitemShut {NoStop}%
\bibitem [{\citenamefont {Singh}\ and\ \citenamefont
  {Han}(2019)}]{PhysRevB.99.174426}%
  \BibitemOpen
  \bibfield  {author} {\bibinfo {author} {\bibfnamefont {V.~K.}\ \bibnamefont
  {Singh}}\ and\ \bibinfo {author} {\bibfnamefont {J.~H.}\ \bibnamefont
  {Han}},\ }\bibfield  {title} {\bibinfo {title} {Application of machine
  learning to two-dimensional dzyaloshinskii-moriya ferromagnets},\ }\href
  {https://doi.org/10.1103/PhysRevB.99.174426} {\bibfield  {journal} {\bibinfo
  {journal} {Phys. Rev. B}\ }\textbf {\bibinfo {volume} {99}},\ \bibinfo
  {pages} {174426} (\bibinfo {year} {2019})}\BibitemShut {NoStop}%
\bibitem [{\citenamefont {Kwon}\ \emph {et~al.}(2019)\citenamefont {Kwon},
  \citenamefont {Kim}, \citenamefont {Lee},\ and\ \citenamefont
  {Won}}]{PhysRevB.99.024423}%
  \BibitemOpen
  \bibfield  {author} {\bibinfo {author} {\bibfnamefont {H.~Y.}\ \bibnamefont
  {Kwon}}, \bibinfo {author} {\bibfnamefont {N.~J.}\ \bibnamefont {Kim}},
  \bibinfo {author} {\bibfnamefont {C.~K.}\ \bibnamefont {Lee}},\ and\ \bibinfo
  {author} {\bibfnamefont {C.}~\bibnamefont {Won}},\ }\bibfield  {title}
  {\bibinfo {title} {Searching magnetic states using an unsupervised machine
  learning algorithm with the heisenberg model},\ }\href
  {https://doi.org/10.1103/PhysRevB.99.024423} {\bibfield  {journal} {\bibinfo
  {journal} {Phys. Rev. B}\ }\textbf {\bibinfo {volume} {99}},\ \bibinfo
  {pages} {024423} (\bibinfo {year} {2019})}\BibitemShut {NoStop}%
\bibitem [{\citenamefont {Lee}\ and\ \citenamefont {Ahn}(2020)}]{Lee2020}%
  \BibitemOpen
  \bibfield  {author} {\bibinfo {author} {\bibfnamefont {W.~S.}\ \bibnamefont
  {Lee}}\ and\ \bibinfo {author} {\bibfnamefont {K.-H.}\ \bibnamefont {Ahn}},\
  }\bibfield  {title} {\bibinfo {title} {Heterogeneous trp channel model of a
  chordotonal neuron might explain drosophila hearing},\ }\href
  {https://doi.org/10.3938/jkps.76.118} {\bibfield  {journal} {\bibinfo
  {journal} {J. Korean Phys. Soc.}\ }\textbf {\bibinfo {volume} {76}},\
  \bibinfo {pages} {118} (\bibinfo {year} {2020})}\BibitemShut {NoStop}%
\bibitem [{\citenamefont {Lee}\ \emph {et~al.}(2021{\natexlab{a}})\citenamefont
  {Lee}, \citenamefont {Jo},\ and\ \citenamefont {Song}}]{Lee2021_1}%
  \BibitemOpen
  \bibfield  {author} {\bibinfo {author} {\bibfnamefont {W.~S.}\ \bibnamefont
  {Lee}}, \bibinfo {author} {\bibfnamefont {J.}~\bibnamefont {Jo}},\ and\
  \bibinfo {author} {\bibfnamefont {T.}~\bibnamefont {Song}},\ }\bibfield
  {title} {\bibinfo {title} {Machine learning for the diagnosis of early-stage
  diabetes using temporal glucose profiles},\ }\href
  {https://doi.org/10.1007/s40042-021-00056-8} {\bibfield  {journal} {\bibinfo
  {journal} {J. Korean Phys. Soc.}\ }\textbf {\bibinfo {volume} {78}},\
  \bibinfo {pages} {373} (\bibinfo {year} {2021}{\natexlab{a}})}\BibitemShut
  {NoStop}%
\bibitem [{\citenamefont {Lee}\ \emph {et~al.}(2021{\natexlab{b}})\citenamefont
  {Lee}, \citenamefont {Kim}, \citenamefont {Cleland},\ and\ \citenamefont
  {Ahn}}]{Lee2021_2}%
  \BibitemOpen
  \bibfield  {author} {\bibinfo {author} {\bibfnamefont {W.~S.}\ \bibnamefont
  {Lee}}, \bibinfo {author} {\bibfnamefont {H.}~\bibnamefont {Kim}}, \bibinfo
  {author} {\bibfnamefont {A.~N.}\ \bibnamefont {Cleland}},\ and\ \bibinfo
  {author} {\bibfnamefont {K.-H.}\ \bibnamefont {Ahn}},\ }\bibfield  {title}
  {\bibinfo {title} {Fast frequency discrimination and phoneme recognition
  using a biomimetic membrane coupled to a neural network},\ }\href
  {https://doi.org/10.1088/1748-3190/abc869} {\bibfield  {journal} {\bibinfo
  {journal} {Bioinspir. Biomim.}\ }\textbf {\bibinfo {volume} {16}},\ \bibinfo
  {pages} {026012} (\bibinfo {year} {2021}{\natexlab{b}})}\BibitemShut
  {NoStop}%
\bibitem [{\citenamefont {Lee}\ and\ \citenamefont
  {Flach}(2020)}]{Lee_SF_2020}%
  \BibitemOpen
  \bibfield  {author} {\bibinfo {author} {\bibfnamefont {W.~S.}\ \bibnamefont
  {Lee}}\ and\ \bibinfo {author} {\bibfnamefont {S.}~\bibnamefont {Flach}},\
  }\bibfield  {title} {\bibinfo {title} {Deep learning of chaos
  classification},\ }\href {https://doi.org/10.1088/2632-2153/abb6d3}
  {\bibfield  {journal} {\bibinfo  {journal} {Mach. Learn.: Sci. Technol.}\
  }\textbf {\bibinfo {volume} {1}},\ \bibinfo {pages} {045019} (\bibinfo {year}
  {2020})}\BibitemShut {NoStop}%
\bibitem [{\citenamefont {Easaw}\ \emph {et~al.}(2023)\citenamefont {Easaw},
  \citenamefont {Lee}, \citenamefont {Lohiya}, \citenamefont {Jalan},\ and\
  \citenamefont {Pradhan}}]{EASAW2023102053}%
  \BibitemOpen
  \bibfield  {author} {\bibinfo {author} {\bibfnamefont {N.}~\bibnamefont
  {Easaw}}, \bibinfo {author} {\bibfnamefont {W.~S.}\ \bibnamefont {Lee}},
  \bibinfo {author} {\bibfnamefont {P.~S.}\ \bibnamefont {Lohiya}}, \bibinfo
  {author} {\bibfnamefont {S.}~\bibnamefont {Jalan}},\ and\ \bibinfo {author}
  {\bibfnamefont {P.}~\bibnamefont {Pradhan}},\ }\bibfield  {title} {\bibinfo
  {title} {Estimation of correlation matrices from limited time series data
  using machine learning},\ }\href
  {https://doi.org/https://doi.org/10.1016/j.jocs.2023.102053} {\bibfield
  {journal} {\bibinfo  {journal} {J. Comput. Sci}\ }\textbf {\bibinfo {volume}
  {71}},\ \bibinfo {pages} {102053} (\bibinfo {year} {2023})}\BibitemShut
  {NoStop}%
\bibitem [{\citenamefont {Miyazaki}(2023)}]{Miyazaki_2023}%
  \BibitemOpen
  \bibfield  {author} {\bibinfo {author} {\bibfnamefont {Y.}~\bibnamefont
  {Miyazaki}},\ }\bibfield  {title} {\bibinfo {title} {Equivariant neural
  networks for spin dynamics simulations of itinerant magnets},\ }\href
  {https://doi.org/10.1088/2632-2153/acffa2} {\bibfield  {journal} {\bibinfo
  {journal} {Machine Learning: Science and Technology}\ }\textbf {\bibinfo
  {volume} {4}},\ \bibinfo {pages} {045006} (\bibinfo {year}
  {2023})}\BibitemShut {NoStop}%
\bibitem [{\citenamefont {Chen}\ \emph {et~al.}(2018)\citenamefont {Chen},
  \citenamefont {Chung}, \citenamefont {Gao}, \citenamefont {Chen},
  \citenamefont {Stone}, \citenamefont {Kolesnikov}, \citenamefont {Huang},\
  and\ \citenamefont {Dai}}]{PhysRevX.8.041028}%
  \BibitemOpen
  \bibfield  {author} {\bibinfo {author} {\bibfnamefont {L.}~\bibnamefont
  {Chen}}, \bibinfo {author} {\bibfnamefont {J.-H.}\ \bibnamefont {Chung}},
  \bibinfo {author} {\bibfnamefont {B.}~\bibnamefont {Gao}}, \bibinfo {author}
  {\bibfnamefont {T.}~\bibnamefont {Chen}}, \bibinfo {author} {\bibfnamefont
  {M.~B.}\ \bibnamefont {Stone}}, \bibinfo {author} {\bibfnamefont {A.~I.}\
  \bibnamefont {Kolesnikov}}, \bibinfo {author} {\bibfnamefont
  {Q.}~\bibnamefont {Huang}},\ and\ \bibinfo {author} {\bibfnamefont
  {P.}~\bibnamefont {Dai}},\ }\bibfield  {title} {\bibinfo {title} {Topological
  spin excitations in honeycomb ferromagnet ${\mathrm{cri}}_{3}$},\ }\href
  {https://doi.org/10.1103/PhysRevX.8.041028} {\bibfield  {journal} {\bibinfo
  {journal} {Phys. Rev. X}\ }\textbf {\bibinfo {volume} {8}},\ \bibinfo {pages}
  {041028} (\bibinfo {year} {2018})}\BibitemShut {NoStop}%
\bibitem [{\citenamefont {Hawkins}(2004)}]{Hawkins2004}%
  \BibitemOpen
  \bibfield  {author} {\bibinfo {author} {\bibfnamefont {D.~M.}\ \bibnamefont
  {Hawkins}},\ }\bibfield  {title} {\bibinfo {title} {The problem of
  overfitting},\ }\href {https://doi.org/10.1021/ci0342472} {\bibfield
  {journal} {\bibinfo  {journal} {J. Chem. Inf. Comput.}\ }\textbf {\bibinfo
  {volume} {44}},\ \bibinfo {pages} {1} (\bibinfo {year} {2004})}\BibitemShut
  {NoStop}%
\bibitem [{\citenamefont {Hendrycks}\ and\ \citenamefont
  {Gimpel}(2023)}]{gelu}%
  \BibitemOpen
  \bibfield  {author} {\bibinfo {author} {\bibfnamefont {D.}~\bibnamefont
  {Hendrycks}}\ and\ \bibinfo {author} {\bibfnamefont {K.}~\bibnamefont
  {Gimpel}},\ }\href@noop {} {\bibinfo {title} {Gaussian error linear units
  (gelus)}} (\bibinfo {year} {2023}),\ \Eprint
  {https://arxiv.org/abs/1606.08415} {arXiv:1606.08415 [cs.LG]} \BibitemShut
  {NoStop}%
\bibitem [{\citenamefont {Jolliffe}(1986)}]{Jolliffe:1986}%
  \BibitemOpen
  \bibfield  {author} {\bibinfo {author} {\bibfnamefont {I.}~\bibnamefont
  {Jolliffe}},\ }\href@noop {} {\emph {\bibinfo {title} {Principal Component
  Analysis}}}\ (\bibinfo  {publisher} {Springer Verlag},\ \bibinfo {year}
  {1986})\BibitemShut {NoStop}%
\bibitem [{\citenamefont {Shalev-Shwartz}\ and\ \citenamefont
  {Ben-David}(2014)}]{10.5555/2621980}%
  \BibitemOpen
  \bibfield  {author} {\bibinfo {author} {\bibfnamefont {S.}~\bibnamefont
  {Shalev-Shwartz}}\ and\ \bibinfo {author} {\bibfnamefont {S.}~\bibnamefont
  {Ben-David}},\ }\href@noop {} {\emph {\bibinfo {title} {Understanding Machine
  Learning: From Theory to Algorithms}}}\ (\bibinfo  {publisher} {Cambridge
  University Press},\ \bibinfo {address} {USA},\ \bibinfo {year}
  {2014})\BibitemShut {NoStop}%
\bibitem [{\citenamefont {Aron}\ \emph {et~al.}(2014)\citenamefont {Aron},
  \citenamefont {Barci}, \citenamefont {Cugliandolo}, \citenamefont {Arenas},\
  and\ \citenamefont {Lozano}}]{Aron_2014}%
  \BibitemOpen
  \bibfield  {author} {\bibinfo {author} {\bibfnamefont {C.}~\bibnamefont
  {Aron}}, \bibinfo {author} {\bibfnamefont {D.~G.}\ \bibnamefont {Barci}},
  \bibinfo {author} {\bibfnamefont {L.~F.}\ \bibnamefont {Cugliandolo}},
  \bibinfo {author} {\bibfnamefont {Z.~G.}\ \bibnamefont {Arenas}},\ and\
  \bibinfo {author} {\bibfnamefont {G.~S.}\ \bibnamefont {Lozano}},\ }\bibfield
   {title} {\bibinfo {title} {Magnetization dynamics: path-integral formalism
  for the stochastic landau–lifshitz–gilbert equation},\ }\href
  {https://doi.org/10.1088/1742-5468/2014/09/P09008} {\bibfield  {journal}
  {\bibinfo  {journal} {J. Stat. Mech.: Theory Exp.}\ }\textbf {\bibinfo
  {volume} {2014}}\bibinfo  {number} { (9)},\ \bibinfo {pages}
  {P09008}}\BibitemShut {NoStop}%
\end{thebibliography}%

\clearpage
\newpage

\appendix

\section{Dataset preparation} \label{app:dataset}

\begin{figure}[t!]
    \centering
   \includegraphics[width=.9\textwidth]{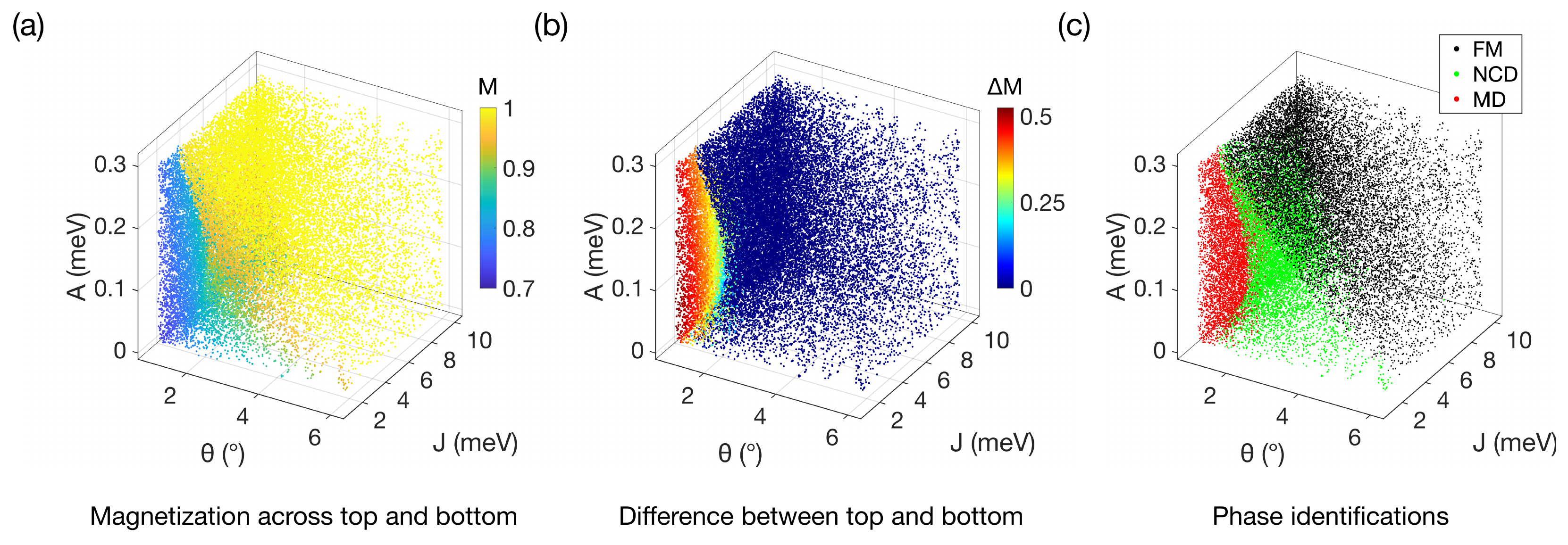}
    \caption[Phase identifications]{Magnetic phase identifications for full datasets obtained in the atomistic spin simulation. (a) Out-of-plane magnetization ($M$) across both top and bottom layers as a function of intralayer exchange coupling ($J$), twist angle ($\theta$), and single-ion anisotropy ($A$). (b) Difference between top and bottom layer magnetizations ($\Delta M$) as a function of the same three parameters. (c) Visual representation for identified magnetic phases as a function of the three parameters. The identification is based on the results on (a–b) and the classification scheme presented in Eq.~\eqref{eq_app:phase_identification}. Black, green, and red markers indicate the FM, NCD, and MD phases, respectively.}
    \label{figS1}
\end{figure}

Our atomistic spin simulations involved 23,000 randomly generated parameter sets ($\theta$, $J$, $A$), as illustrated in Fig.~\ref{figS1}. We computed the magnetic ground state for each parameter set by employing 50 different random initial spin configurations and relaxing each configuration using an iterative optimization method \cite{Kim2024}. During the relaxation, periodic boundary conditions were imposed on the boundary of the moiré supercell. The magnetic ground state is determined by selecting the lowest energy configuration among the resulting energy-minimized spin configurations.

To classify the magnetic ground states obtained in the simulation, we calculated the following order parameters:
\begin{subequations} \label{eq_app:order_parameters}
\begin{align}
    M & = \frac{1}{2}(M_z^\textrm{top}+M_z^\textrm{bot}) \label{eq_app:M} \\ 
    \Delta M & = |M_z^\textrm{top} - M_z^\textrm{bot}|. \label{eq_app:DeltaM}
\end{align}
\end{subequations}
Here, $M_z^\textrm{top}$ and $M_z^\textrm{bot}$ represent the averaged out-of-plane magnetization in the top and bottom layers, respectively. They are defined as:
\begin{align}
M_z^\textrm{top} & = \frac{1}{N_\textrm{top}} \sum_{i\in\textrm{top}} n_{i}^{z},  \nonumber \\ 
M_z^\textrm{bot} & = \frac{1}{N_\textrm{bottom}} \sum_{i\in\textrm{bottom}} n_{i}^{z},
\end{align}
Here, $n_{i}^{z}$ denotes out-of-plane components of a normalized vector $\bm{n}_i = (n_{i}^{x}, n_{i}^{y}, n_{i}^{z})$ for spin at $i$-site, $N_\textrm{top}$ and $N_\textrm{bottom}$ denote the numbers of spins on the top and bottom layers, respectively. The calculation results for $M$ and $\Delta M$ are presented in Fig.~\ref{figS1}(a–b). The magnetic ground states are categorized based on the following criteria:
\begin{align} \label{eq_app:phase_identification}
    & \textrm{FM Phase}:  ~~M = 1, ~~ \Delta M = 0, \nonumber \\ 
    & \textrm{NCD Phase}: M < 1, ~~ \Delta M = 0, \nonumber \\
    & \textrm{MD Phase}:  ~~M < 1, ~~ \Delta M > 0.
\end{align}
The classification of the magnetic ground states using Eq.~\eqref{eq_app:phase_identification} is presented in Fig.~\ref{figS1}(c)

Utilizing this result, we consistently omitted FM states from our dataset. Consequently, we prepared 17,292 magnetic ground states within the NCD and MD phases, which were utilized to generate 17,292 paired images in our dataset. In the process of generating images, the supercell of moiré superlattices is first repeated and then tailored to a fixed length dimension of 100 nm. This regularization ensures uniformity across moiré superlattices of different sizes. The generated images have dimensions of (200, 200).

\section{Neural network architectures} \label{app:network_structures}

\begin{figure}[t!]
    \centering
   \includegraphics[width=0.98\textwidth]{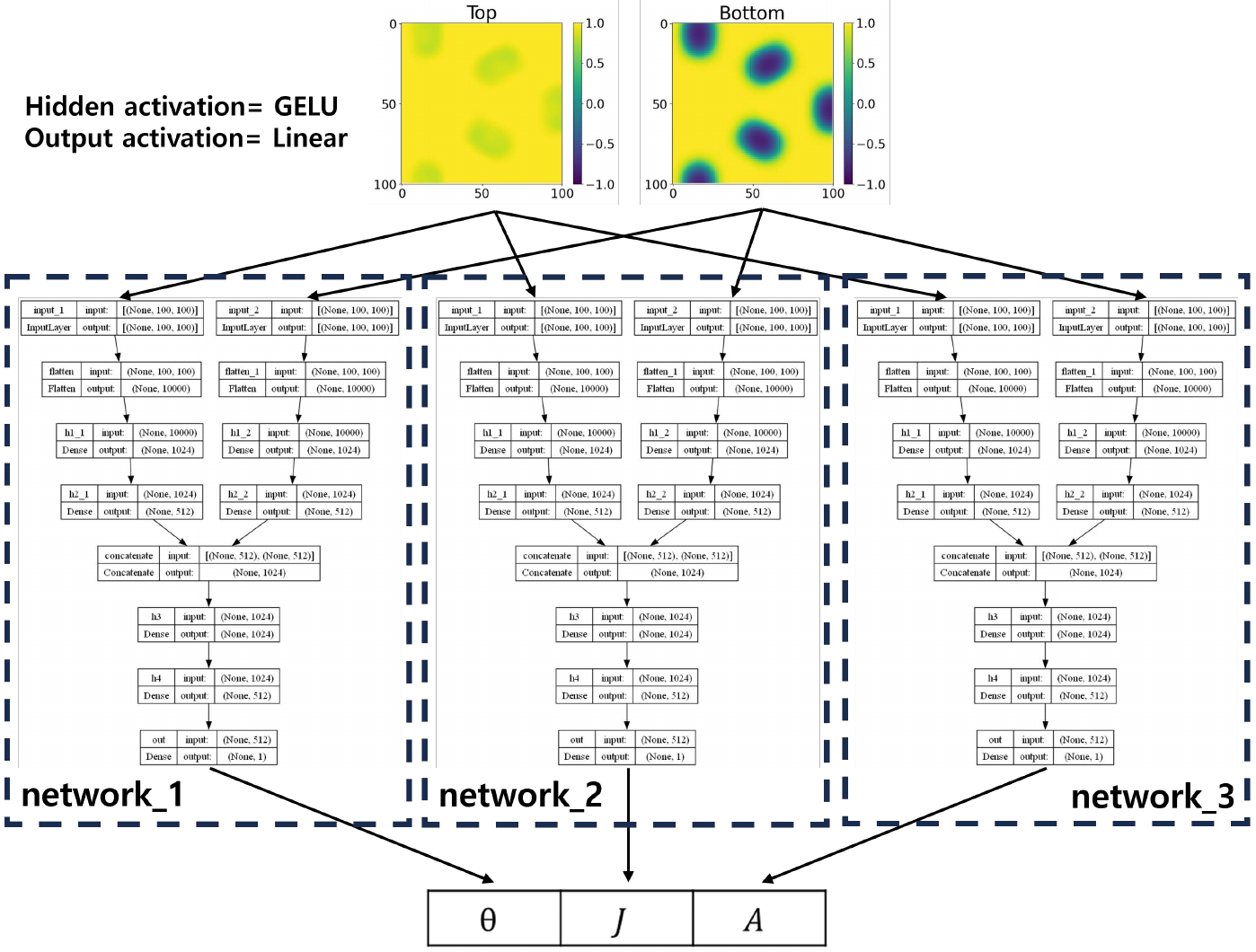}
    \caption{Neural network structure of the regression model. The model takes two pixel arrays (‘Top’ and ‘Bottom’) as input and produces three parameters $(\theta, J, A)$ as output. The model comprises three independent networks (denoted as \texttt{`network\_1'}, \texttt{`network\_2'}, \texttt{`network\_3'}) for regressing the three parameters. Each network has two input layers (\texttt{`input\_1'} and \texttt{`input\_2'}), two flattening layers (\texttt{`flatten'} and \texttt{`flatten\_1'}), and four hidden layers (\texttt{`h1\_1'}, \texttt{`h2\_1'} and \texttt{`h1\_2'}, \texttt{`h2\_2'}), with two dedicated to each input. Additionally, it has a concatenating layer (\texttt{`concatenate'}), two more hidden layers (\texttt{`h3'} and \texttt{`h4'}) for the concatenated data, and an output layer (\texttt{`out'}).}
    \label{figS2}
\end{figure}

Figure~\ref{figS2} provides a detailed illustration of the neural network structure employed in the regression model. The model incorporates three neural networks (labeled as \texttt{`network\_1'}, \texttt{`network\_2'}, \texttt{`network\_3'}) for regressing three parameters $(\theta, J, A)$ from given two pixel arrays (`Top True' and `Bottom True'). These arrays are derived from the first quadrant of full images sized at (200, 200), obtained in the simulations. This preprocessing step ensures efficiency in processing while maintaining consistency due to the four-fold rotational symmetry inherent in each image. Each network accepts the two arrays through dedicated input layers (\texttt{`input\_1'} and \texttt{`input\_2'}), flattening them into two lists (\texttt{`concatenate'}). Subsequently, the two lists undergo independent processing through two consecutive hidden layers (\texttt{`h1\_1'}, \texttt{`h2\_1'} and \texttt{`h1\_2'}, \texttt{`h2\_2'} for each list), ultimately reducing to 512 features each, combined into a single list sized at 1024 (\texttt{`concatenate'}). This list then undergoes processing through two consecutive hidden layers (\texttt{`h3'}, \texttt{`h4'}), resulting in a further reduction to 512 features. The output layer (\texttt{`out'}) of the neural network regresses these features to predict a single parameter from the set $(\theta, J, A)$.

\begin{figure}[t!]
    \centering
   \includegraphics[width=0.8\textwidth]{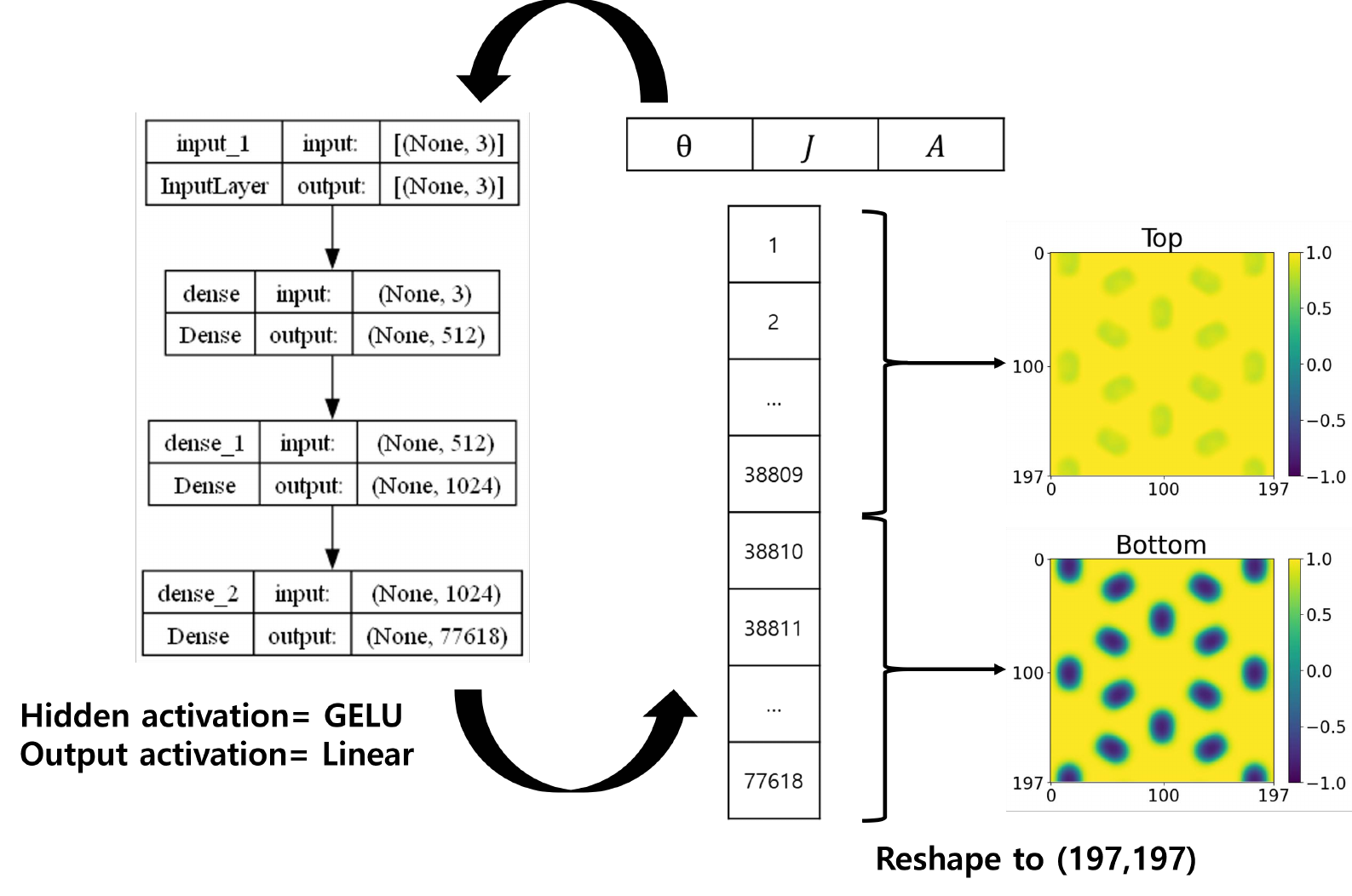}
    \caption{Neural network structure of the generative model. The generative model takes three parameters $(\theta, J, A)$ as input and produces two pixel arrays (labeled as ‘Top’ and ‘Bottom’) as output. The network consists of an input layer (denoted as \texttt{`input\_1'}), two hidden layers (\texttt{`dense'} and \texttt{`dense\_1'}), and an output layer (\texttt{`dense\_2'}).}
    \label{figS3}
\end{figure}

Figure~\ref{figS3} provides a detailed illustration of the neural network structure employed in the generative model. The input layer (labeled as \texttt{`input\_1'}) processes a list of three parameters $(\theta, J, A)$. This input undergoes further processing through two hidden layers (\texttt{`dense'} and \texttt{`dense\_1'}) and the output layer (\texttt{`dense\_2'}). The result is a list of 77,618 floating-point numbers representing as many features. Subsequently, this output list is reshaped into two pixel arrays with dimensions (197, 197), serving as the predicted domain array images (‘Top Pred.’ and ‘Bottom Pred.’).

\section{Validation of the regression model against thermal noises} \label{app:thermal_noises}

\begin{figure}[t]
    \centering
   \includegraphics[width=0.85\textwidth]{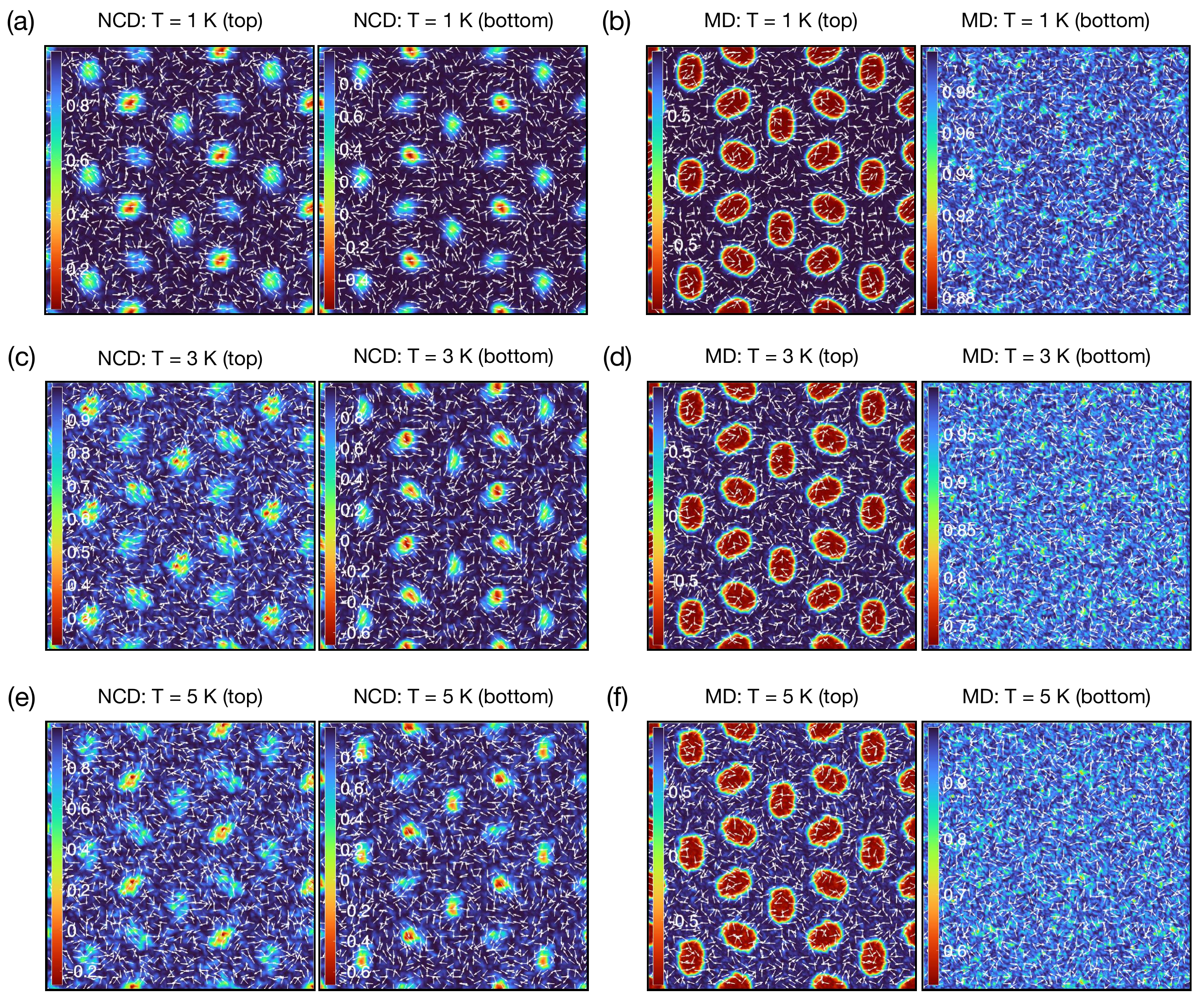}
    \caption{Magnetic configurations at various temperatures, simulated using by the s-LLG equation. (a,c,e) NCD configurations corresponding to Fig.~\ref{fig2}(d) at $T=1$ K, $T=3$ K, and $T=5$ K, respectively. (b,d,f) MD configurations corresponding to Fig.~\ref{fig2}(e) at $T=1$ K, $T=3$ K, and $T=5$ K, respectively.  Left and right panels in each plot correspond to the top and bottom layers, respectively. The color scale depicts the out-of-plane magnetization directions, and the arrows indicate the in-plane magnetization directions.}
    \label{figS4}
\end{figure}

We generated noisy test sets with thermal fluctuations at finite temperatures by solving the stochastic Landau–Lifshitz–Gilbert (s-LLG) equation:
\begin{align}
    \frac{d\bm{m}_{i}}{dt} = -\gamma \bm{m}_{i} \times \bm{H}_{\textrm{eff},i} + \alpha \bm{m}_{i} \times \frac{d\bm{m}_{i}}{dt}.
\end{align}
In this equation, \(\alpha\) is the dimensionless Gilbert damping coefficient, representing energy dissipation due to environmental interactions. \(\gamma = 1.76 \times 10^{11} \, \text{rad}/(\text{second} \cdot \text{tesla})\) is the gyromagnetic ratio. The magnetization vector \(\bm{m}_{i}\) is defined by:
\begin{align}
    \bm{m}_{i} = \frac{\bm{\mu}_{i}}{|\bm{\mu}_{i}|},
\end{align}
where \(\bm{\mu}_{i} = -\gamma \bm{S}_{i}\) is the magnetic moment of the spin \(\bm{S}_{i}\). The effective magnetic field \(H_{\textrm{eff},i}^{\alpha}(t)\) is defined as:
\begin{align}
    H_{\textrm{eff},i}^{\alpha}(t) = \frac{1}{\gamma} \sum_{j} \sum_{\beta} J_{ij}^{\alpha\beta} S_{j}^{\beta} + H_{\textrm{ran},i}(t),
\end{align}
where the magnetic interaction term \(J_{ij}^{\alpha\beta}\) is given by:
\begin{align}
    J_{ij}^{\alpha\beta} = -J \delta_{\langle i, j\rangle} \delta_{\alpha, \beta} - \delta_{l(i), l(j)} \delta_{\alpha, z} \delta_{\beta, z} A + J_{ij}^{\perp} \delta_{l(i), t} \delta_{l(j), b} \delta_{\alpha, \beta}.
\end{align}
Here, \(\delta_{\langle i, j\rangle}\) indicates that \(i\) and \(j\) are nearest neighbors in the same layer, while \( \delta_{l(i), t} \delta_{l(j), b}\) signifies that \(i\) and \(j\) belong to the top (\(t\)) and bottom (\(b\)) layers, respectively. The random thermal field \(H_{\textrm{ran},i}(t)\) represents the effect of thermal noise, with a white-noise autocorrelation function:
\begin{align}
    \left\langle H_{\textrm{ran},i}(t) H_{\textrm{ran},j}(t') \right\rangle = 2D \delta_{i, j} \delta(t - t'),
\end{align}
where the fluctuation amplitude \(D\) is given by:
\begin{align}
    D = \frac{\alpha}{\gamma |\bm{\mu}_{i}|} k_B T.
\end{align}
The random field \(H_{\textrm{ran},i}(t)\) is generated using a Gaussian random number generator with a variance of \(2D/dt\), with temperatures ranging from 1 to 5 kelvins. For the simulations, we used the original test sets as initial configurations and then computed their time evolution based on the stochastic Landau-Lifshitz-Gilbert (s-LLG) equation. In these calculations, we set \(\alpha = 0.01\) and used a time step of 10 femtoseconds, integrating the s-LLG equation for a duration of 100 picoseconds to ensure that the spin configurations reached equilibrium. Figure~\ref{figS4} shows examples of the noisy magnetic configurations obtained from these calculations.

\begin{figure}[t]
    \centering
    \includegraphics[width=0.85\textwidth]{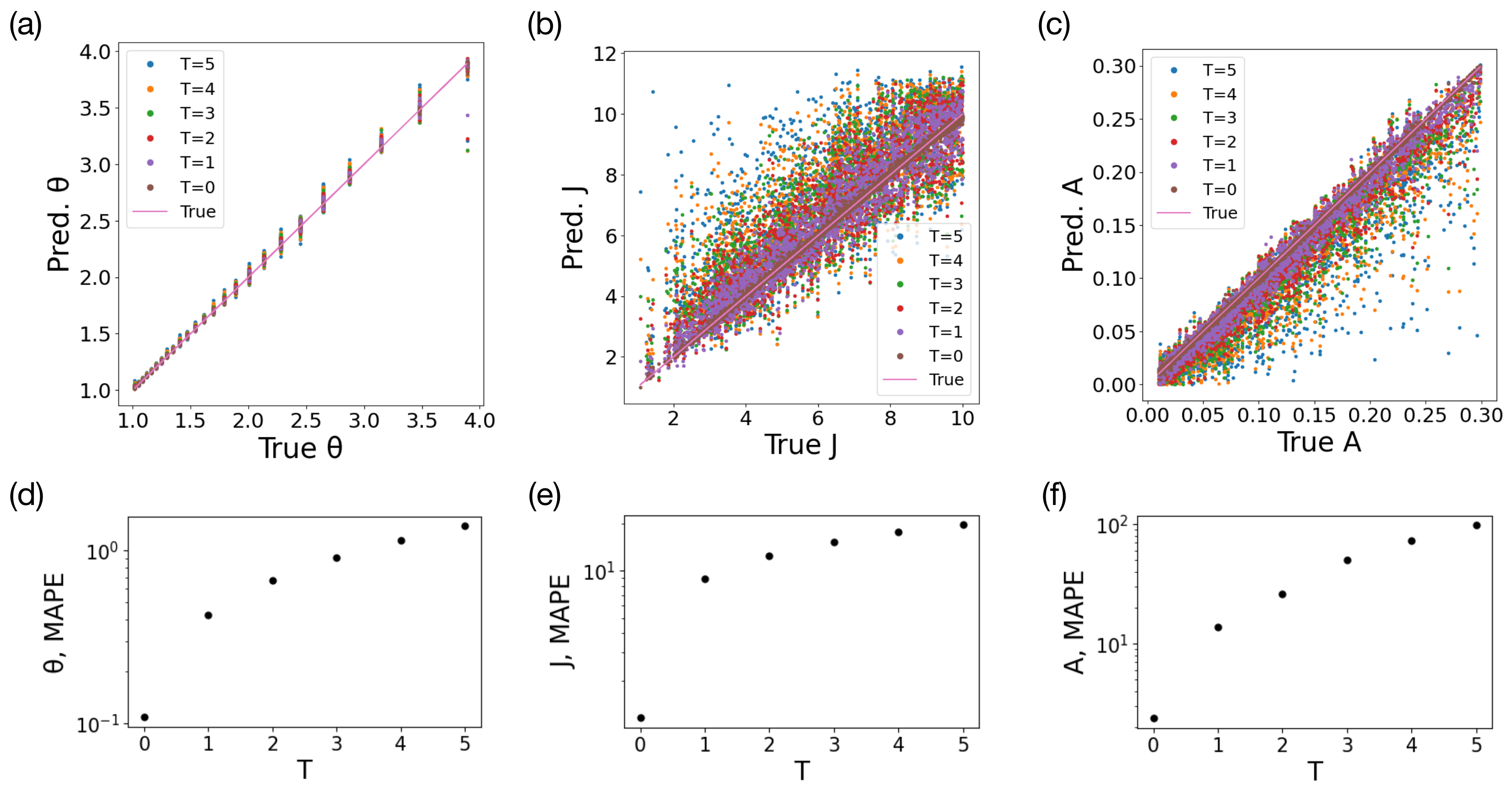}
    \caption{Performance of the trained networks in the regression model amidst thermal noises, specifically within the NCD phase. (a–c) Parameter estimation results derived from noisy test sets in comparison to the noiseless original test set (denoted by $T=0$). Here, $T$ denotes the temperature in the kelvin unit. The $x$-axis denotes the true values used in the simulation, while the $y$-axis represents the model's estimated values. The pink lines ($y=x$, denoted by True) represent the ideal predictions. (d–f) MAPE values for the estimated parameters at different temperatures ($T$). The markers denote MAPE values for each parameter.}
    \label{figS5}
\end{figure}

\begin{figure}[t]
    \centering
    \includegraphics[width=0.85\textwidth]{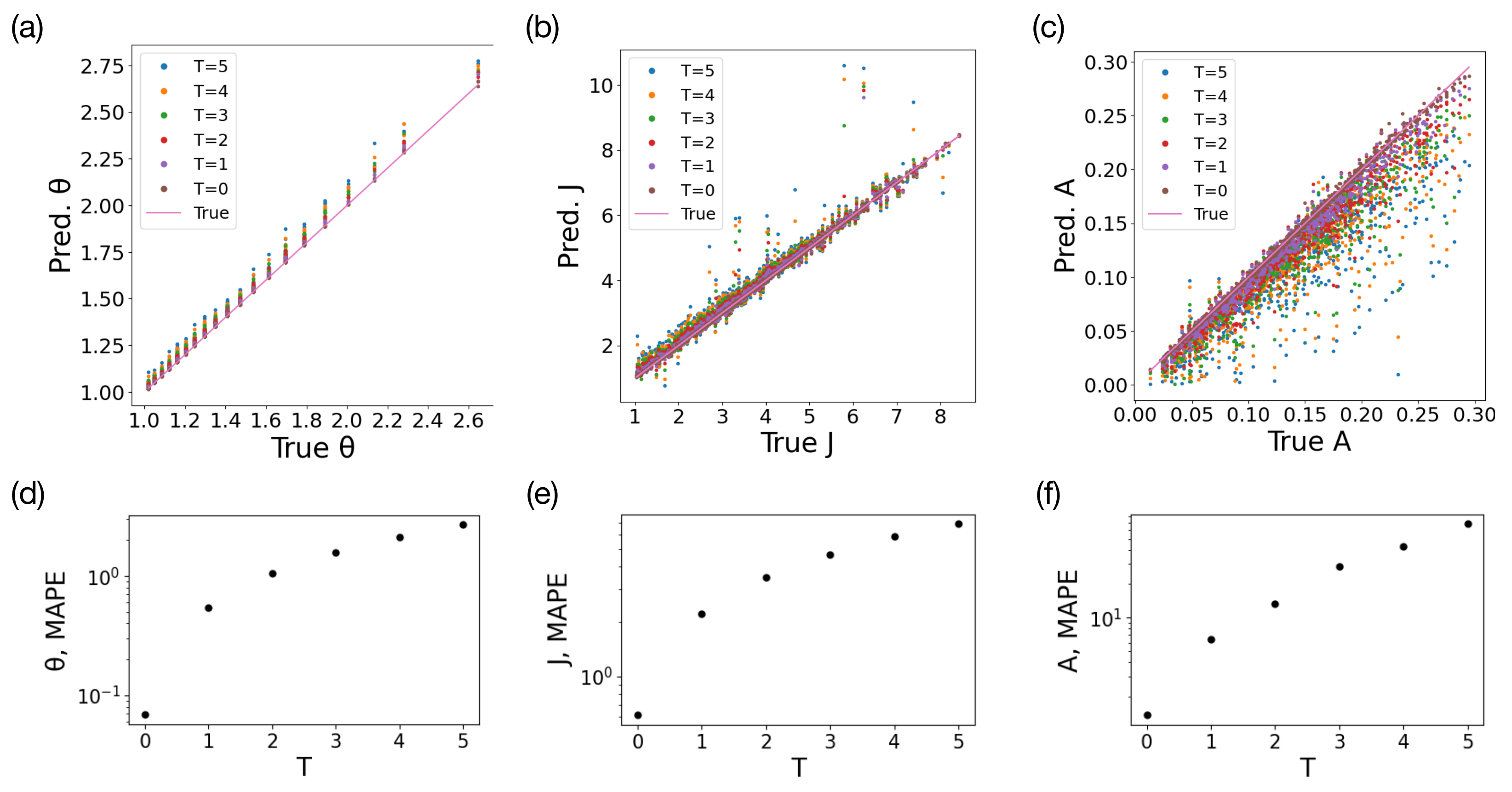}
    \caption{Performance of the trained networks in the regression model amidst thermal noises, specifically within the MD phase. (a–c) Parameter estimation results derived from noisy test sets in comparison to the noiseless original test set (denoted by $T=0$). Here, $T$ denotes the temperature in the kelvin unit. The $x$-axis denotes the true values used in the simulation, while the $y$-axis represents the model's estimated values. The pink lines ($y=x$, denoted by True) represent the ideal predictions. (d–f) MAPE values for the estimated parameters at different temperatures ($T$). The markers denote MAPE values for each parameter.}
    \label{figS6}
\end{figure}

Figures~\ref{figS5} and \ref{figS6} present the parameter estimation results on noisy test sets obtained from the networks trained on the noiseless training set. Our analysis reveals that in both the NCD and MD phases, the trained network maintains its highly accurate predictive capabilities for estimating $\theta$, showing an MAPE value of less than 3\% up to the highest investigated temperature of 5 K (Fig.~\ref{figS5}(d) and Fig.~\ref{figS6}(d)). The network shows robust performance in estimating $J$, with MAPE values of approximately 7\% and 20\% at the highest temperature in the MD and NCD phases, respectively (Fig.~\ref{figS5}(e) and Fig.~\ref{figS6}(e)). The relatively poor performance in the NCD phase might be attributed to the vulnerable domain structure against thermal noises observed in our simulations (see Fig.~\ref{figS4}(a,c,e) and Fig.~\ref{figS4}(b,d,f)), making it challenging for the network to extract this parameter accurately.

Our analysis reveals significant challenges in estimating $A$ from the noisy test sets. For instance, the MAPE values reach approximately 50\% and 30\% at an intermediate temperature of 3 K in the NCD and MD phases, respectively (Fig.~\ref{figS5}(f) and Fig.~\ref{figS6}(f)). This poor predictive capability for $A$ may stem from the substantial impact of the random thermal field, with an estimated energy scale of approximately 0.88 meV at 3 K, significantly stronger than the parameter value of $A \lesssim 0.3$ meV. This suggests that the network struggles to accurately estimate $A$ due to the overwhelming influence of thermal fluctuations at this temperature in determining the detailed morphology of magnetic structures.

\end{document}